\documentclass[universe,article,accept,pdftex,moreauthors]{Definitions/mdpi} 

\firstpage{1} 
\pubvolume{1}
\issuenum{1}
\articlenumber{0}
\pubyear{2025}
\copyrightyear{2025}
\externaleditor{Firstname Lastname}
\datereceived{29 November 2024} 
\daterevised{4 January 2025} 
\dateaccepted{6 January 2025} 
\datepublished{ } 
\hreflink{https://doi.org/} 

\usepackage{color}

\def\comment#1{}

\Title{Distinguishing Compact Objects in Extreme-Mass-Ratio Inspirals by Gravitational Waves}

\TitleCitation{Distinguishing Compact Objects in Extreme-Mass-Ratio Inspirals by Gravitational Waves}

\Author{Lujia Xu 
 $^{1,2}$, Shucheng Yang $^{2,}$*, Wenbiao Han $^{2,3,4,5,}$*, Xingyu Zhong $^{3}$, Rundong Tang $^{2,4}$ and Yuanhao Zhang $^{3}$}

\AuthorNames{Lujia Xu, Shucheng Yang,  Wenbiao Han, Xingyu Zhong, Rundong Tang and Yuanhao Zhang}

\AuthorCitation{Xu, L.; Yang, S.; Han, W.; Zhong, X.; Tang, R.; Zhang, Y.}

\address{%
$^{1}$ \quad School of Physics and Astronomy, University of Glasgow, Glasgow G12 8QQ, UK; 
\\
$^{2}$ \quad Shanghai Astronomical Observatory, Chinese Academy of Sciences, Shanghai 200030, China; 
\\
$^{3}$ \quad Hangzhou Institute for Advanced Study, University of Chinese Academy of Sciences, \linebreak Hangzhou 310124, China; 
\\
$^{4}$ \quad School of Astronomy and Space Science, University of Chinese Academy of Sciences, Beijing 100049, China\\
$^{5}$ \quad Taiji Laboratory for Gravitational Wave Universe (Beijing/Hangzhou), University of Chinese Academy of Sciences, Beijing 100049, China}

\corres{Correspondence: ysc@shao.ac.cn (S.Y.); wbhan@shao.ac.cn (W.H.)}

\abstract{Extreme-mass-ratio inspirals (EMRIs) are promising gravitational-wave (GW) sources for space-based GW detectors. EMRI signals typically have long durations, ranging from several months to several years, necessitating highly accurate GW signal templates for detection. In most waveform models, compact objects in EMRIs are treated as test particles without accounting for their spin, mass quadrupole, or tidal deformation. In this study, we simulate GW signals from EMRIs by incorporating the spin and mass quadrupole moments of the compact objects. We evaluate the accuracy of parameter estimation for these simulated waveforms using the Fisher Information Matrix (FIM) and find that the spin, tidal-induced quadruple, and spin-induced quadruple can all be measured with precision ranging from $10^{-2}$ to $10^{-1}$, particularly for a mass ratio of $\sim$$10^{-4}$. Assuming the ``true'' GW signals originate from an extended body inspiraling into a supermassive black hole, we compute the signal-to-noise ratio (SNR) and Bayes factors between a test-particle waveform template and our model, which includes the spin and quadrupole of the compact object. Our results show that the spin of compact objects can produce detectable deviations in the waveforms across all object types, while tidal-induced quadrupoles are only significant for white dwarfs, especially in cases approaching an intermediate-mass ratio. Spin-induced quadrupoles, however, have negligible effects on the waveforms. Therefore, our findings suggest that it is possible to distinguish primordial black holes from white dwarfs, and, under certain conditions, neutron stars can also be differentiated from primordial black holes.}


\begin{document}


\section{Introduction}
The observations of GWs from compact binary coalescences since 2015 have ushered in a new era of astronomy \cite{ligo2016observation, ligo2017gw170817}. There are abundant sources that emit GWs in the low-frequency band, which can be observed by future space-borne GW detectors such as the Laser Interferometer Space Antenna (LISA) \cite{amaro2007intermediate}. EMRIs, which consist of central massive black holes (MBHs) and orbiting compact objects, are important GW sources for space-borne GW detectors \cite{amaro2007intermediate, gair2010lisa}. 

{The small object in an EMRI could be a stellar-mass black hole, a neutron star, a white dwarf, or another compact object. EMRIs emit gravitational waves (GWs) as the secondary objects orbit the central massive black holes (MBHs). Furthermore, in certain scenarios, such as when the compact objects interact with the matter in the MBH accretion disc, EMRIs can also produce high-energy electromagnetic (EM) emissions \cite{Kejriwal_2024}. Such emissions typically arise from cases where the compact objects are either captured by the accretion disc in an active galactic nucleus (AGN) and migrate inwards or form directly within the disc and evolve as they move toward the MBH. Such EMRI sources, capable of emitting both GWs and EM signals, hold significant potential for multi-messenger detections. However, this study focuses exclusively on gravitational wave signals. In this context, understanding the dynamical behavior of compact objects in EMRIs is crucial for advancing low-frequency GW astronomy.}

Presently, the spin interaction of relativistic systems has become an important subject. For this reason, it is a real concern to properly understand the dynamics of extended bodies in curved space–time that includes classical spin. The {dynamics} is simple when one considers the point-particle approximation. Nevertheless, once considering the structure, the problem is hard to solve. In Newtonian mechanics, a solution to the problem of the motion of $N$ isolated bodies with internal structure was first proposed by F. Tisserand \cite{tisserand1889traite}. In his work, Tisserand was able to separate the external and internal motion of the body by considering the linearity of the equations. In this way, it was possible to describe the dynamics of one of the bodies with high precision. However, in contrast to Newtonian gravity, the field equations of general relativity are coupled and nonlinear. Therefore, it is not possible to apply the same methods as in Newtonian mechanics. In general relativity, it is well known that a point-particle follows a geodesic. However, when considering extended bodies, it is necessary to take into account the effect of the body in the space–time metric, an effect known as the self-field \cite{wald2009introduction}. 

The first approach to solving the extended body problem in general relativity goes back to 1937 with the work of M. Mathisson \cite{mathisson1937neue}, who demonstrated the existence of an interaction between the Riemann curvature tensor and the spin of the moving particle in the equations of motion. {Mathisson} showed that it is possible to define force, center-of-mass, torque, and mass in a relativistic theory. The problem of extended bodies in general relativity was also considered by Papapetrou \cite{papapetrou1951spinning,honl1940innere,corinaldesi1951spinning}, where he uses a similar approach. Later, B. Tulczyjew and W. Tulzcyjew improved and developed the methods of {Mathisson} \cite{tulczyjew1959motion, tulzcyjew1962recent}. On the other hand, improvements in the definition of the center-of-mass were made by Moller and others in refs. \cite{moller1949definition,beiglbock1967center,dixon1964covariant,dixon1970dynamics,dixon1973definition,ehlers1977dynamics}. Today, the equations that describe the motion of extended bodies with spin and mass are known as the {Mathisson-Papapetrou}-Dixon equations (MPD). 

When considering the inspiral orbital motion of an equal-mass spinning binary system, it is crucial to consider the higher-order multipole moment contributions \cite{singh2015the}. Nevertheless, in the case of EMRIs, it makes sense to truncate the multipole expansion and focus mainly on the pole-dipole approximation. In general, calculations involving spinning objects with dimensions sufficiently small compared to the background space–time's local curvature radius can be performed with good approximation by employing the MPD equations of motion. In the literature, there are a variety of astrophysical situations where the MPD equations are used to show the impact of spin-curvature interactions between spinning particles and black holes \cite{wald1972gravitational, tod1976spinning, semerak1999spinning, suzuki1998innermost}. From the numerical point of view, it is possible to investigate the limits of stability for the MPD equations \cite{suzuki1998innermost, suzuki1999signature, hartl2003dynamics, hartl2003survey}. In the work of S. Suzuki and K. Maeda \cite{suzuki1998innermost}, the authors studied the stability of circular orbits for spinning test particles in Kerr space–time. They showed that orbits in the radial direction are stable, while some circular orbits become unstable in the direction perpendicular to the equatorial plane. Moreover, in the case of particles with higher spin, the innermost stable circular orbit (ISCO) appears before the minimum of the effective potential in the equatorial plane disappears. 

Using the MPD equation, it is possible to derive predictions about the generation of gravitational waves that are expected to occur from spin-induced deviations away from geodesic motion. In ref. \cite{mino1996gravitational}, Yasushi Mino et al. used Teukolsky, Sasaki, and Nakamura's formalisms to perturb the Kerr black hole and calculate the energy flux and the waveform induced by a spinning particle falling from infinity into a rotating black hole. Due to the combination of Teukolsky formalism with the MPD equations, the authors found two additional effects related to the particles' spin: the first effect is due to the spin-spin interaction force, and the other is due to the contribution of the energy-momentum tensor of the spinning particle. According to the numerical calculations, the authors argue that these effects are significant. In this sense, a deeper understanding of the relativistic two-body problem requires accurate and general results from both numerical and analytic computations.

Theoretically, it is possible to recognize the compact objects by their quadrupolar deformation in EMRIs by GWs \cite{rahman2021prospect}, and so is the spin of compact objects in EMRIs. While this is controversial for spin, some researchers believe that the spin of compact objects in EMRIs is not observable \cite{piovano2021assessing, huerta2011importance}. In the present work, we use the MPD equation to simulate the GW signals of EMRIs that consider the compact objects' spin and quadrupole to study to what extent these parameters could influence the GW signals. {It is worth noting that before conducting a series of calculations and simulations, we clarify that environmental effects, such as hydrodrag from interactions between compact objects and an accretion disc \cite{narayan2003magnetically}, for simplicity, are not considered in this study.} We find that for the gravitational wave (GW) signals of EMRIs, both the spin and tidal effects of compact objects are influential, particularly when the compact objects are white dwarfs. However, spin-induced quadrupoles have no significant impact, even in EMRIs with small mass ratios. Our results also demonstrate that primordial black holes (PBHs) with sub-solar masses can be clearly distinguished from white dwarfs. In certain cases, neutron stars can be differentiated from PBHs, specifically when the neutron star spin exceeds that of the PBHs. Moreover, if we replace the MPD equation with the test particle approximation in the waveform templates, the matched-filtering SNRs of GW signals, assuming really from extended bodies, remain almost unchanged. Therefore, for the aim of detecting EMRIs, we may omit the spin and quadrupole of the compact object in constructing the waveform templates of EMRIs. This will greatly reduce the parameter space and the computation cost for searching EMRIs in the data. 

This paper is organized as follows: Section \ref{EOM} starts with the equation of motion for extended bodies. Then, we introduce the GW waveforms for EMRIs In Section \ref{GWofEMRI}. In Section \ref{results}, we present our results in detail. Finally, we conclude Section \ref{conclusion}.  Through this paper,  we use the Einstein summation convention.

\section{Characteristics of Compact Objects in EMRIs}
\label{EOM}
The accuracy of the waveform templates is crucial in GW detection. Therefore, we need to calculate accurate orbits of compact objects. The MPD equations describe the motion of extended bodies in curved space–time with spin and mass multipole moments. The higher-order terms of the MPD equations (expanded by multipole moments) show that the inner structure of compact objects slightly influences the orbit \cite{lincoln1990coalescing}. In the present work, we considered the EMRI model that uses extended bodies with spin and quadrupole moments, which fit the actual orbits better than the models that use test particles. In this case, the motion equations of the compact objects are (using natural units of $G = c = 1$) \cite{dixon1970dynamics}
\begin{align}
    \dot p^{\mu}=-\frac{1}{2}S^{\alpha\beta}v^{\rho}R^{\mu}_{ \rho\alpha\beta}-\mathcal{F}^\mu,
\end{align}
\begin{align}
    \dot S^{\alpha\beta}=2p^{[\alpha}v^{\beta]}+G^{\alpha\beta},
\end{align}
where $p^{\mu}$ is the four-momentum of small compact objects, defined as $p^{\mu}=mu^{\mu}$, and $m$ is defined as the dynamical mass of compact objects, satisfying the condition $m^{2}=-p^{\mu}p_{\mu}$, which depends on the four-momentum of compact objects, so $m$ is not a constant. The dot means the differential concerning proper time $\tau$. $u^\mu$ is the dynamical velocity of bodies, satisfying the condition $u^{\mu}u_{\mu}=-1$. $S^{\alpha\beta}$ is a second-order anti-symmetrical spin tensor, which satisfies the spin conservation condition $S^{2}=S^{\mu}S_{\mu}=\frac{1}{2}S^{\mu\nu}S_{\mu\nu}$. The kinematical four-velocity of extended bodies $v^{\rho}=dz^{\rho}/d\tau$, and $z(\tau)$ is the world line of the extended bodies' mass center, which is determined by the supplementary condition $u_{\lambda}S^{\kappa\lambda}=0$. $R^{\mu}_{ \rho\alpha\beta}$ is the Riemannian curvature tensor, $\mathcal{F}^{\mu}$ and $G^{\alpha\beta}$ are the coupling terms between the quadrupole and background gravitational field:
\begin{align}
\mathcal{F}^{\mu}=\frac{1}{6}J^{\alpha\beta\gamma\sigma}{\nabla}^{\mu}R_{\alpha\beta\gamma\sigma},
\end{align}
\begin{align}
G^{\alpha\beta}=\frac{4}{3}J^{\gamma\delta\epsilon[\alpha}R_{ \delta\epsilon\gamma}^{\beta]},
\end{align}
where $J^{\gamma\delta\epsilon\sigma}$ is the mass quadrupole tensor that has the same symmetry as $R_{\alpha\beta\gamma\sigma}$. The relationship between the four-velocity and the four-momentum is \cite{han2017dynamics}
\begin{align}
    m^{2}v^{\sigma}=mp^{\sigma}-\mathcal{F}^{\sigma\rho}p_{\rho}+\frac{2mp^{\rho}R_{\mu\rho\alpha\beta}S^{\sigma\mu}S^{\alpha\beta}-2p_{\delta}\mathcal{F}^{\rho\delta}R_{\mu\rho\alpha\beta}S^{\sigma\mu}S^{\alpha\beta} + 4m^{2}\mathcal{F}_{\mu}S^{\sigma\mu}}{4m^{2}+R_{\mu\rho\alpha\beta}S^{\sigma\mu}S^{\alpha\beta}}.
\end{align}
When 
 $v^{\mu}v_{\mu}=-1$, $\tau$ is the proper time. Generally speaking, the kinematical mass $\overline m=p^{\mu}v_{\mu}$ is not equal to the dynamical mass $m$, but in the present work we use the orthogonal condition $u^{\mu}v_{\mu}=-1$, where $m=\overline m$. 

The mass quadrupole tensor takes the form \cite{steinhoff2012influence}:
\begin{align}
    J^{\alpha\beta\gamma\delta}=\frac{3m}{\overline m^3}p^{[\alpha}Q^{\beta][\gamma}p^{\delta]},
\end{align}
where the quadrupole of extended bodies is given by
\begin{align}
    Q^{\alpha\beta}=C_{Q}S^{\alpha}_{\lambda}S^{\beta\lambda}+\frac{1}{\overline m^{2}}\mu_2R^{\alpha\beta\gamma\delta}u_{\gamma}u_{\delta},
\end{align}
where $C_{\rm Q}$ is a dimensionless constant to measure the spin-induced quadrupole and is related to the equation of state (EOS) of extended bodies. Providing the radius and mass of a rotating compact object, $C_{\rm Q}$ can be approximately expressed by the equation \cite{laarakkers1999quadrupole}:
\begin{align}
C_{\rm Q} \approx -\frac{25}{8}\frac{R c^2}{Gm},
\label{eqn:cq}
\end{align}
where $G$ is the gravitational constant,  $c$ is the speed of light in a vacuum, and $R$ is the body's radius. $C_{\rm Q}$ varies for different compact objects. For black holes, $|C_{\rm Q}| = 1$; for neutron stars, $|C_{\rm Q}|$ varies from 2 to 20 in different EOS \cite{uchikata2016tidal, narikawa2021gravitational, saleem2022population}; for white dwarfs, $|C_{\rm Q}|$ takes about $10^4$ \cite{boshkayev2012general}, and $\mu_{2}$ represents the quadrupole produced by the tidal effect, and it takes the form \cite{steinhoff2012influence}
\begin{align}
\mu_2=\frac{2k_2}{3}(R/\frac{Gm}{c^2})^5\nu^4,
\label{eqn:mu2}
\end{align}
where $k_2$ is a dimensionless tidal love parameter determined by the EOS of compact objects. $\nu$ is the symmetric mass ratio of EMRI ($\nu = mM / (m + M)^2$). Binnington and Poisson \cite{binnington2009relativistic} proposed a relativistic tidal parameter theory that applies to compact objects with strong inner gravity, for black holes $k_2=0$; for neutron stars $k_2\sim 0.1$; for white dwarfs $k_2 \sim 0.01$ \cite{taylor2020love}. According to Equation \eqref{eqn:mu2}, for black holes, $\mu_{2}=0$; for neutron stars with radius of 10--20 km, and mass of $1 - 2 \mathrm{M_{\odot}}$, ${\mu_{2} / \nu^{4} }$ is $\sim$$10^{2}$ to $\sim$$10^{3}$. for white dwarfs, as show in Table \ref{tab:table1}, we calculate some values of $C_\mathrm{Q}$ and ${\mu_{2} / \nu^{4} }$ for white dwarfs. 

\begin{table}[H] 
\caption{$C_\mathrm{Q}$ and ${\mu_{2} / \nu^{4} }$ of white dwarfs in different mass and radius.
\label{tab:table1}}
\newcolumntype{C}{>{\centering\arraybackslash}X}
\begin{tabularx}{\textwidth}{CCCC}
\toprule
\textbf{Mass (}\boldmath{$\mathrm{M_{\odot}}$}\textbf{)}	& \textbf{Radius (}\boldmath{$\mathrm{R_{\odot}}$}\textbf{)} & \boldmath{$|C_{Q}|$} & \boldmath{$\mu_{2}/ \nu^{4}$} \\
\midrule
$0.75$  & $0.009$  &17,668 & $3.85\times 10^{16}$  \\
$1.003$ & $0.0084$ &12,330& $6.38\times 10^{15}$ \\
$1.1$   & $0.0031$ &$4149$ & $2.75\times 10^{13}$ \\
$1.2$   & $0.0055$ &$6748$ & $3.13\times 10^{14}$  \\
$1.28$  & $0.0041$ &$4716$ & $5.22\times 10^{13}$  \\
$1.3$   & $0.004$  &$4530$ & $4.27\times 10^{13}$  \\
$1.33$  & $0.003$  &$3321$ & $9.04\times 10^{12}$  \\
\bottomrule
\end{tabularx}
\end{table}

The spin angular momentum of the compact object $s$ is another important parameter in this work. For a stellar black hole, the maximum spin angular momentum $s = m^2$ (using natural units of G = c = 1). However, for primordial black holes, research shows that PBHs possess negligible spin at formation \cite{sasaki2016primordial, de2019initial}, and baryonic accretion can spin up primordial black holes at masses larger than $\sim$$10 \mathrm{M_{\odot}}$ \cite{de2020evolution}. Neutron stars and white dwarfs can have spin magnitudes a little larger than $m^2$. For convenience, we use dimensionless $\hat{s} = s / m^2$ in the following parts of this article, and for the calculation of orbits, we use the spin parameter (using natural units of G = c = 1)
\begin{align}
\label{eqn:s}
S = \frac{s}{mM} \approx \hat{s} \nu.
\end{align}

The 
 $S$ of compact objects in EMRIs is much less than one. As shown in Table \ref{tab:table2}, we summarize several physical characteristics of the small compact object in EMRI \cite{maggiore2018gravitational}, such as stellar origin black holes(SOBHs), PBHs, neutron stars(NSs), and white dwarfs(WDs).

\begin{table}[H] 
\caption{Physical characteristics of compact objects.
\label{tab:table2}}
\begin{adjustwidth}{-\extralength}{0cm}
\newcolumntype{C}{>{\centering\arraybackslash}X}
\begin{tabularx}{\fulllength}{CCCCCC}
\toprule
&\textbf{Mass (}\boldmath{$\mathrm{M_{\odot}}$}\textbf{)}	& \boldmath{$\hat{s}$}	& \boldmath{$k_{2}$} & \boldmath{$\mu_{2}/ \nu^{4} $} & \boldmath{$|C_{Q}|$}\\
\midrule
SOBH&2$\sim$5--50$\sim$150          & 0--1 & $0$  & $0$  & $1$\\
PBH &  $\sim$$10^{-19}$--$\sim$$10^{3}$   & $\sim$0\textsuperscript{1} & $0$ & $0$  & $1$\\
NS  &1.1--2.1                     & $\lesssim $1.3 & $\sim $0.1 & $\sim$$10^{2}$--$\sim$$10^{3}$ & 2--20\\
WD  &0.2--1.44                     & $\lesssim $10 & $\sim $0.01 & $\sim$$10^{12}$--$\sim$$10^{16}$ & $\sim$$10^{3}$--$\sim$$10^{4}$\\
\bottomrule
\end{tabularx}
\end{adjustwidth}
\noindent{\footnotesize{\textsuperscript{1} The specific spin values of PBHs still remain debatable. Since PBH spins reflect their cosmological origin, several different scenarios have been proposed for the formation of PBHs. One of the earliest PBH formation mechanisms was due to Chiba and Yokoyama \cite{chiba2017spin}, who found that PBHs can be generated in the early universe through the collapse of sufficiently density perturbations and gave theoretical predictions of the probability distribution of PBH spins, indicating that the PBHs tend to have low spins with $ a_{s}\lesssim 0.4$ at the radiation-dominated stage. In contrast, Flores et al. \cite{flores2021spins}. discussed the PBH spins produced in different cosmological scenarios, one of which is an intermediate matter-dominated epoch, where they found that PBHs formed from the merger of particles or scalar fields in the absence of radiative cooling can have a large range of spins. Recently, using the probability distribution of the spin amplitude of PBH spins given by Luca et al. \cite{de2019initial,de2020evolution} who computed it in a more sophisticated way based on the standard results of peak theory, Chongchitan \cite{chongchitnan2021extreme} statistically quantified the rarity of extreme-spin PBHs, which means the existence of PBHs with spin exceeding astrophysical limits, such as the Thorne limit $a_{s}=0.998$, and they found that roughly one in a million PBHs was formed with spin {$a_{s} \gtrsim 0.8$} and one in a hundred million formed with spin exceeding the Thorne limit. Thus, considering different cosmological scenarios, we take a range of $\hat{s}\sim (0,1)$ for the PBH spins in order to provide a more comprehensive approach to our work in distinguishing compact objects.}}
\end{table}

When setting the orbit configuration and calculating the orbits of EMRIs, we consider the innermost stable circular orbit (ISCO) and tidal radius as the restrictions. ISCO is the smallest stable orbit for a test particle orbiting a massive object. For a rotating BH, the radius of ISCO is as follows:
\begin{align}
\label{eqn:r_isco}
R_{\mathrm{ISCO}} = M[3 + Z_{2} \pm \sqrt{ (3 - Z_{1})(3 + Z_{1} + 2 Z_{2})}],
\end{align}
where  
\begin{align}
\label{eqn:z1}
Z_{1} = 1 + (1-a^2)^{1/3}[(1+a)^{1/3} + (1-a)^{1/3}],
\end{align}
\begin{align}
\label{eqn:z2}
Z_{2} = \sqrt{3a^2 + Z_{1}^{2}}.
\end{align}
with $a$ as the rotation parameter of the rotating black hole. Equation \eqref{eqn:r_isco} takes a negative sign when the orbit is prograde and takes a positive sign when the orbit is retrograde. For an EMRI, inside the ISCO, no stable circular orbits exist. Another important concept is tidal radius \cite{bardeen1972rotating}
\begin{align}
\label{eqn:r_tidal}
R_{\mathrm{tidal}} = 2^{1/3} \frac{R}{m} \nu ^{2/3}M.
\end{align}

\begin{figure}[H]
\includegraphics[width=12cm]{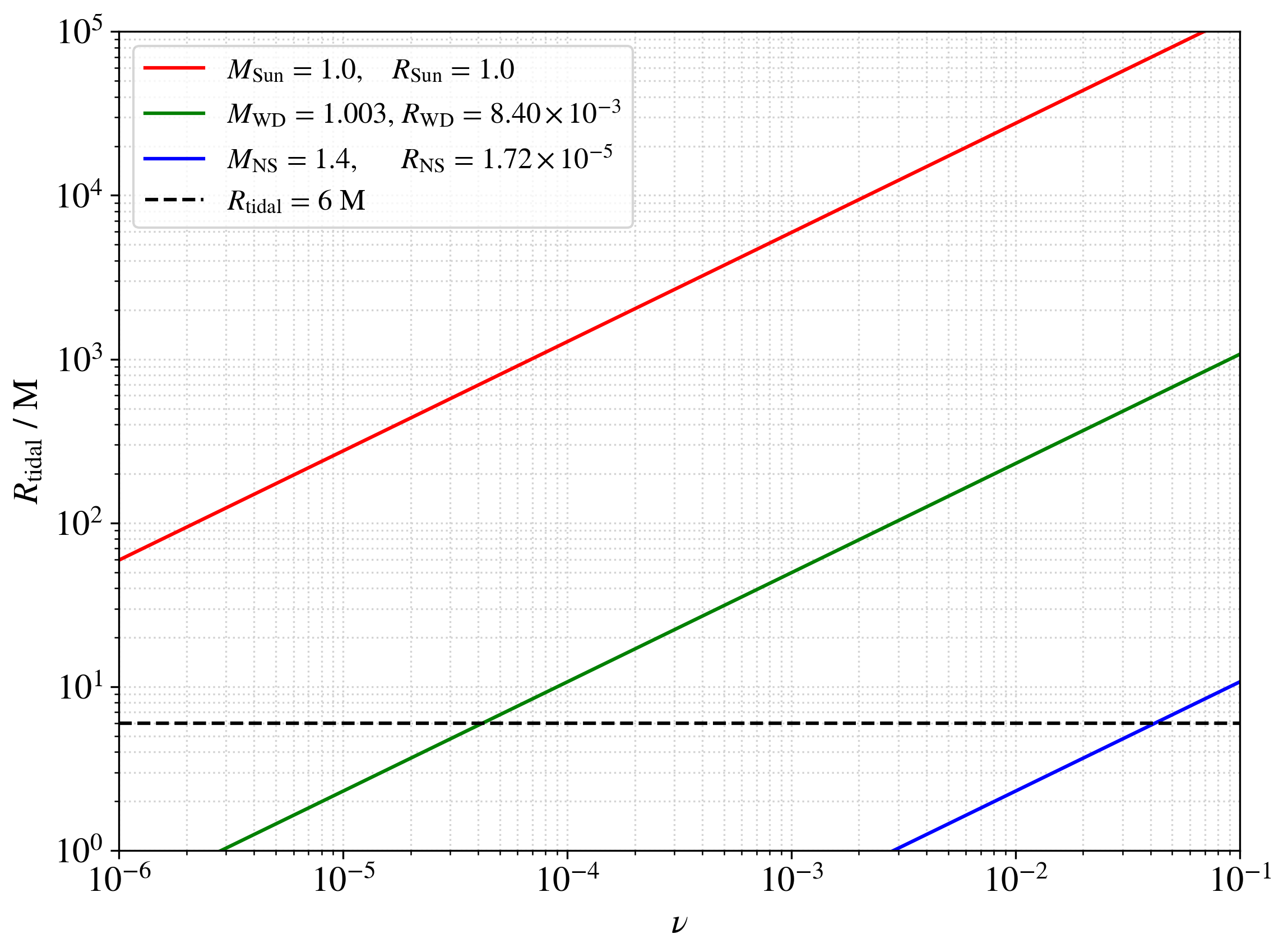}
\caption{The tidal radius $R_{\mathrm{tidal}}$ of EMRIs varies with different $\nu$ for each orbiting body. Specifically, the red curve represents the Sun, the green curve represents a white dwarf, and the blue curve represents a neutron star. Additionally, the black dotted curve corresponds to $R_{\rm tidal}$ of $6 \mathrm{M}$, which equals the $R_{\mathrm{ISCO}}$ for a non-rotating black hole.
\label{fig:r_tidal}}
\end{figure}

Inside the tidal radius, some compact objects, such as white dwarfs and neutron stars, would be torn apart by the tidal force of the central black hole. According to Equation \eqref{eqn:r_tidal},  Figure \ref{fig:r_tidal} shows the tidal radius of EMRIs for different $\nu$ for different orbiting bodies (the sun, a white dwarf, and a neutron star). For an EMRI with $\nu = 10^{-6}$, the tidal radius for the sun is around $100 \mathrm{M}$, and for white dwarf and neutron stars are less than $1  \mathrm{M}$. for an EMRI with $\nu = 10^{-4}$, the tidal radius for the sun is $1278.88  \mathrm{M}$, the tidal radius for white dwarf is $10.71  \mathrm{M}$, the tidal radius for neutron star is $0.107  \mathrm{M}$. 
\section{The Gravitational Waves Signals of EMRIs}
\label{GWofEMRI}
We first integrate the MPD equation to derive the trajectories and then calculate GW signals. There are several conserved quantities for the MPD equations in Kerr space–time, i.e., the energy $E$ and the $z$ component of angular momentum $L_z$ \cite{han2017dynamics}
\begin{align}
    E=-p_t+\frac{1}{2}g_{t\mu,\nu}S^{\mu\nu},
\end{align}
\begin{align}
    L_z=p_{\phi}-\frac{1}{2}g_{\phi\mu,\nu}S^{\mu\nu}.
\end{align}
In 
 the case of the test particle, the Carter constant $Q$ should be another conserved quantity. While the Carter constant does not exist for an extended body endowed with both spin and quadrupole. Fortunately, due to the tiny spin parameter $S$ in EMRIs, it is possible to find an approximate ``Carter constant'' at first order \cite{rudiger1981conserved, rudiger1983conserved}, so we can still assume the following relationship:
\begin{align}
    Q=L_z^{2}\tan^{2}\iota,
\end{align}
where $Q$ is the approximate ``Carter constant'' $\iota$ is the inclination angle of the orbit and can be obtained by
\begin{align}
    \iota=\frac{\pi}{2}-\theta_{\rm min},
\end{align}
where $\theta_{\rm min}$ is the minimum of the $\theta$ coordinate along the trajectory. The semi-latus rectum $p$ and eccentricity $e$ can be calculated by \cite{gair2006improved}
\begin{align}
    p=\frac{2r_ar_p}{r_a+r_p}, e=\frac{r_a-r_p}{r_a+r_p},
\end{align}
where $r_p, r_a$ is the periapsis and apoapsis of the EMRI orbits. Then, the EMRI orbits are parameterized by $e$,$p$,$\iota$.

When considering the effect of radiation reaction, we neglected the time derivative of $S^{\mu\nu}$ because it will involve a higher order of mass ratio in the calculation, and the radiation reaction $F^{\mu}$ can be recovered from the adiabatic radiation fluxes \cite{xin2019gravitational}
\begin{align} 
    & \dot Eu^t = -g_{tt}F^t-g_{t\phi}F^\phi, \label{fluxes1}\\
    & \dot L_zu^t = g_{t\phi}F^t+g_{\phi\phi}F^\phi, \\
    & \dot Qu^t = 2g^{2}_{\theta\theta}u^{\theta}F^{\theta}+2\cos^{2}{\theta}a^{2}E\dot E+2\cos^{2}{\theta}\frac{L_z\dot L_z}{\sin^{2}{\theta}}, \\
    & g_{\mu\nu}u^{\mu}F^{\nu} = 0, \label{fluxes4}
\end{align}

For the fluxes of $E,  L_z$, and $Q$, Glampedakis, Hughes, and Kennefick (GHK) \cite{glampedakis2002approximating} used the lowest order Post-Newtonian (PN) fluxes which are modified from Ryan \cite{ryan1996effect}, and proposed a scheme for computing approximate generic EMRI trajectories, then constructed the inspirals by evolving $E$, $L^{z}$ and $Q$.  It is not accurate enough to evolve inspirals. To adopt higher order PN fluxes to the GHK scheme and ensure accuracy, ref. \cite{gair2006improved} rewrote the fluxes in terms of $p$ and $e$ and included higher order terms in $e$. In particular, we must have the factor $(1-e^{2})^{\frac{3}{2}}$ to ensure the behavior is qualitatively correct in the high eccentricity case. For generic orbits, following \cite{gair2006improved}  we have
\begin{adjustwidth}{-\extralength}{0cm}
\begin{equation}
\begin{split}
&\left(\dot{E}\right) = -\frac{32}{5}\nu(\frac{1}{\tilde{p}})^{5}(1-e^{2})^{3/2}[g_{1}(e)-a(\frac{1}{\tilde {p}})^{3/2}g_{2}(e)\cos{\iota}-(\frac{1}{\tilde{p}})g_{3}(e)+\pi(\frac{1}{\tilde{p}})^{3/2}g_{4}(e)-(\frac{1}{\tilde{p}})^{2}g_{5}(e)+a^{2}(\frac{1}{\tilde{p}})^{2}g_{6}(e)\\& -\frac{527}{96}a^{2}(\frac{1}{\tilde{p}})^{2}\sin^{2}{\iota}],\\
&\left(\dot{L_{z}}\right)= -\frac{32}{5}\nu(\frac{1}{\tilde{p}})^{7/2}(1-e^{2})^{3/2}[g_{9}(e)\cos{\iota}+a(\frac{1}{\tilde{p}})^{3/2}\{g^{a}_{10}(e)-\cos^{2}{\iota}g^b_{10}(e)\}-(\frac{1}{\tilde{p}})g_{11}(e)\cos{\iota}+\pi(\frac{1}{\tilde{p}})^{3/2}g_{12}(e)\cos{\iota}\\&-(\frac{1}{\tilde{p}})^{2}g_{13}(e)\cos{\iota}+a^{2}(\frac{1}{\tilde{p}})^{2}\cos{\iota}(g_{14}(e)-\frac{45}{8}\sin^{2}{\iota})].
\end{split}
\end{equation}
\end{adjustwidth}

To describe the orbits with high inclination, i.e., $\iota\approx\pi/2$ and avoid discontinuous transition across the pole, evolving $Q$ instead of $\iota$ is necessary \cite{gair2006improved}. A better expression for $\dot{Q}$ is derived from the high PN angular momentum flux $(2, 2)$ so that we can ensure $\dot{Q}$ is finite at the pole:
\vspace{-12pt}\begin{adjustwidth}{-\extralength}{0cm}
\begin{equation}
\begin{split}
 &\left(\dot{Q}\right) = -\frac{64}{5}\nu(\frac{1}{\tilde{p}})^{7/2}\sqrt{Q}\sin{\iota}(1-e^{2})^{3/2}[g_9(e)-q(\frac{1}{\tilde{p}})^{3/2}\cos{\iota}g_{10}^b(e)-(\frac{1}{\tilde{p}})g_{11}(e) + \pi(\frac{1}{\tilde{p}})^{3/2}g_{12}(e)\\&-(\frac{1}{\tilde{p}})^{2}g_{13}(e)+q^{2}(\frac{1}{\tilde{p}})^{2}(g_{14}(e)-\frac{45}{8}\sin^{2}{\iota})],
\end{split}
\end{equation}
\end{adjustwidth}
where $q=a/M, \nu=\mu/M, \tilde{p}=p/M$ and the expressions of coefficients $g(e)$ are \cite{gair2006improved}
\vspace{-12pt}\begin{adjustwidth}{-\extralength}{0cm}
\begin{equation}
\begin{split}
g_{1}\left(e\right) &=1+\frac{73}{24}e^{2}+\frac{37}{96}e^{4}, g_{2}\left(e\right) \frac{73}{12}+\frac{823}{24}e^{2}+\frac{946}{32}e^{4}+\frac{491}{192}e^{6}, g_{3}\left(e\right) =\frac{1247}{336}+\frac{9181}{672}e^{2},\\
g_{4}\left(e\right) &=4+\frac{1375}{48}e^{2}, g_{5}\left(e\right) =\frac{44711}{9072}+\frac{172157}{2592}e^{2}, g_{6}\left(e\right) =\frac{33}{16}+\frac{359}{32}e^{2},\\
g_{7}\left(e\right) &=\frac{8191}{672}+\frac{44531}{336}e^{2}, 
g_{8}\left(e\right) =\frac{3749}{336}-\frac{5143}{168}e^{2}, 
g_{9}\left(e\right) =1+\frac{7}{8}e^{2},\\
g_{10}\left(e\right)&=\frac{61}{12}+\frac{119}{8}e^{2}+\frac{183}{32}e^{4}, g_{11}\left(e\right) =\frac{1247}{336}+\frac{425}{336}e^{2}, g_{12}\left(e\right)=4+\frac{97}{8}e^{2},\\
g_{13}\left(e\right)&=\frac{44711}{9072}+\frac{302893}{6048}e^{2}, 
g_{14}\left(e\right) =\frac{33}{16}+\frac{95}{16}e^{2}, g_{15}\left(e\right) =\frac{8191}{672}+\frac{48361}{1344}e^{2},\\
g_{16}\left(e\right) &=\frac{417}{56}-\frac{37241}{672}e^{2}, g_{10}^{a}\left(e\right) =\frac{61}{24}+\frac{63}{8}e^{2}+\frac{95}{64}e^{4}, g_{10}^{b}\left(e\right)=\frac{61}{8}+\frac{91}{4}e^{2}+\frac{461}{64}e^{4},\\
\end{split}
\end{equation}
\end{adjustwidth}

With $\dot{E}, \dot{L_z}$ and $\dot{Q}$ at hand, we can obtain the radiation reactions from Equations \eqref{fluxes1}--\eqref{fluxes4}
\vspace{-12pt}\begin{adjustwidth}{-\extralength}{0cm}
\begin{equation}
\begin{split}
    F^t=\frac{(g_{\phi\phi}\dot{E}+g_{t\phi}\dot{L_{z}})u^t}{g_{t\phi}^2-g_{tt}g_{t\phi}}, F^{\phi}=\frac{(g_{t\phi}\dot{E}+g_{tt}\dot{L_{z}})u^t}{g_{tt}g_{t\phi}-g_{t\phi}^2}, F^{\theta}=\frac{\dot{Q}u^t-2\cos^{2}{\theta}a^2E\dot{E}-2\cos^{2}{\theta}\frac{L_z\dot L_z}{\sin^{2}{\theta}}}{2g_{\theta\theta}^2u^{\theta}},\\
    F^r=-\frac{1}{g_{rr}u^r}(\frac{\dot{Q}u^t-2\cos^{2}{\theta}a^2E\dot{E}-2\cos^{2}{\theta}\frac{L_z\dot L_z}{\sin^{2}{\theta}}}{2g_{\theta\theta}}+\frac{(g_{tt}g_{\phi\phi}-g_{t\phi}^2)\dot{L_z}u^tu^{\phi}}{g_{tt}g_{t\phi}-g_{t\phi}^2}+\frac{(g_{t\phi}^2-g_{tt}g_{\phi\phi})\dot{E}u^tu^t}{g_{tt}g_{t\phi}-g_{t\phi}^2}).
\end{split}
\end{equation}
\end{adjustwidth}

Now 
 we can rewrite the MPD equation with the radiation reactions
\begin{align}
    \dot p^{\mu}=-\frac{1}{2}S^{\alpha\beta}v^{\rho}R^{\mu}_{ \rho\alpha\beta}-\mathcal{F}^\mu + F^\mu\,.
\end{align}
In 
 this way, the orbits of the compact objects will evolve under gravitational radiation with the above equation. Note that here, we use the test particle's fluxes and relationships to calculate the radiation reactions. Due to the extreme mass ratio, the influence of spin and quadrupole will be at the second order of mass ratio, which can be omitted here.
After obtaining the orbit, we calculate the gravitational waveform of EMRIs by the quadrupole approximation \cite{babak2007kludge}
\begin{align}
    \overline{h}^{jk}(t,{\rm x})=\frac{2}{r}\left[\ddot{I}^{jk}(t')\right]_{t'=t-r},
\end{align}
\begin{align}
    I^{jk}=\mu x'^{j}_{p}x'^{k}_{p},
\end{align}
where $\overline{h}^{\mu\nu}=h^{\mu\nu}-\frac{1}{2}\eta^{\mu\nu}\eta^{\rho\sigma}h_{\rho\sigma}$ is the trace-reversed metric perturbation \cite{xin2019gravitational}. We transform the waveform into a transverse–traceless gauge, and we obtain the plus and cross components of the waveform observed at latitudinal angle $\Theta$ and azimuthal angle $\Phi$:
\vspace{-12pt}\begin{adjustwidth}{-\extralength}{0cm}
\begin{align}
\label{eqn:waveform}
h_{+}&=h^{\Theta\Theta}-h^{\Phi\Phi}\\
&=[\cos^{2}{\Theta}(h^{xx}\cos^{2}{\Phi}+h^{xy}\sin{2\Phi}h^{yy}\sin^{2}{\Phi})+h^{zz}\sin^{2}{\Theta}-\sin2{\Theta}(h^{xz}\cos{\Phi}+h^{yz}\cos{\Phi})]\notag\\
&-(h^{xx}\sin^{2}{\Phi}-h^{xy}\sin2{\Phi}+h^{yy}\cos{\Phi})\notag,\\
h_{\times}&=2h^{\Theta\Phi}\\
&=2[\cos{\Theta}(-\frac{1}{2}h^{xx}\sin2{\Phi}+h^{xy}\cos2{\Phi}+\frac{1}{2}h^{yy}\sin2{\Phi})+\sin{\Theta}(h^{xz}\sin{\Phi}-h^{yz}\cos{\Phi})]\notag.
\end{align}
\end{adjustwidth}

\section{Data Analysis and Results}
\label{results}
As shown in Table \ref{tab:table2}, the compact object within the mass range 1--1.44 $\rm M_{\odot}$ could be a PBH, a neutron star, or a white dwarf. If we could obtain the spin or quadrupole moment information for the compact object in the EMRI,  we may distinguish its constitution. The following calculations set the mass of compact objects to $1 \rm M_{\odot}$ for convenience. Then, we focus on the spin and tidal effects and analyze if we can recognize the small objects using GW signals from EMRIs.

\begin{figure}[H]
\begin{adjustwidth}{-\extralength}{0cm}
\centering
\includegraphics[width=18.5cm]{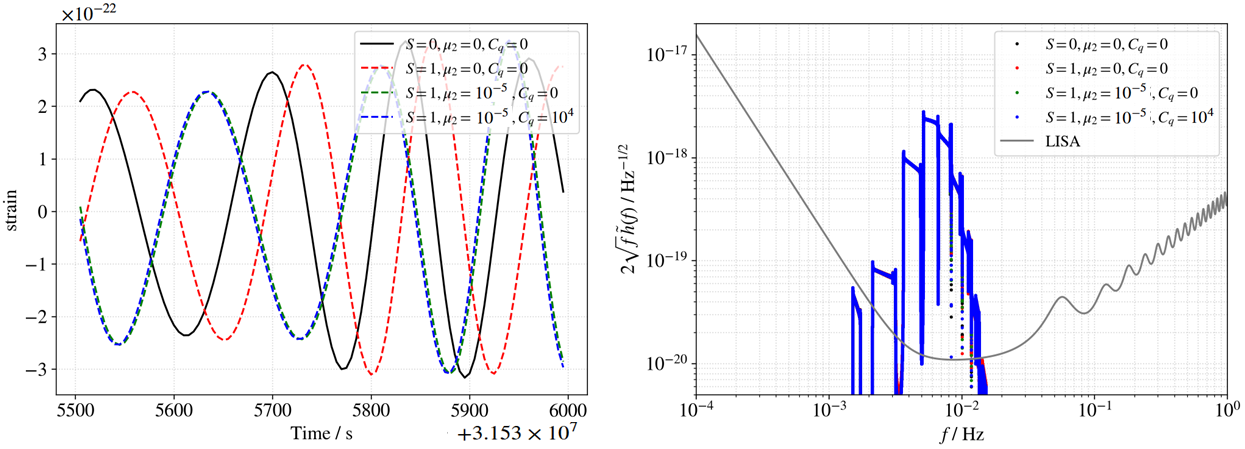}
\end{adjustwidth}
\caption{The 
 GW signals ($h_{+}$) of EMRIs in different configurations for $\hat{s}$, $\mu_{2}$ and $C_{Q}$. The duration of signals is one year, $M_{\mathrm{bh}} = 10^{6} \mathrm{M_{\odot}}$, $M_{\mathrm{co}} = 1.0 \mathrm{M_{\odot}}$, $a = 0.9$, $\iota_{0} = 0.0$, $e  = 0.2$ and $p = 5$. The (\textbf{left}) panel displays the GW signals in the time domain, and the (\textbf{right}) panel displays them in the frequency domain. In the frequency domain, the grey curve represents the sensitivity curve of LISA.
\label{fig:wave}}
\end{figure}

In Figure \ref{fig:wave}, we plot GW signals($h_{+}$) of different EMRI configurations in the time domain(left panel) and frequency domain(right panel). In the frequency domain, the grey curve represents the sensitivity curve of LISA. We show the influence on waveform phase by spin $\hat{s}$, tidal-induced quadrupole $\mu_{2}$, and spin-induced quadrupole $C_{\rm Q}$ of compact objects. The duration of signals is one year, the mass of central black hole $M_{\mathrm{bh}} = 10^{6} \mathrm{M_{\odot}}$, the mass of compact object $M_{\mathrm{co}} = 1.0 \mathrm{M_{\odot}}$, the Kerr parameter $a = 0.9$, the inclination angle of orbit $\iota_{0} = 0.0$, the eccentricity $e  = 0.2$, the semi latus rectum $p = 5$. The black curve is the initial GW signal without the influence of $\hat{s}$, $\mu_{2}$, and $C_{\rm Q}$. The influence of $\hat{s}$, $\mu_{2}$, and $C_{\rm Q}$ are added successively in the red, green, and blue curves.

\begin{figure}[H]
\hspace{-1em}\includegraphics[width=12cm]{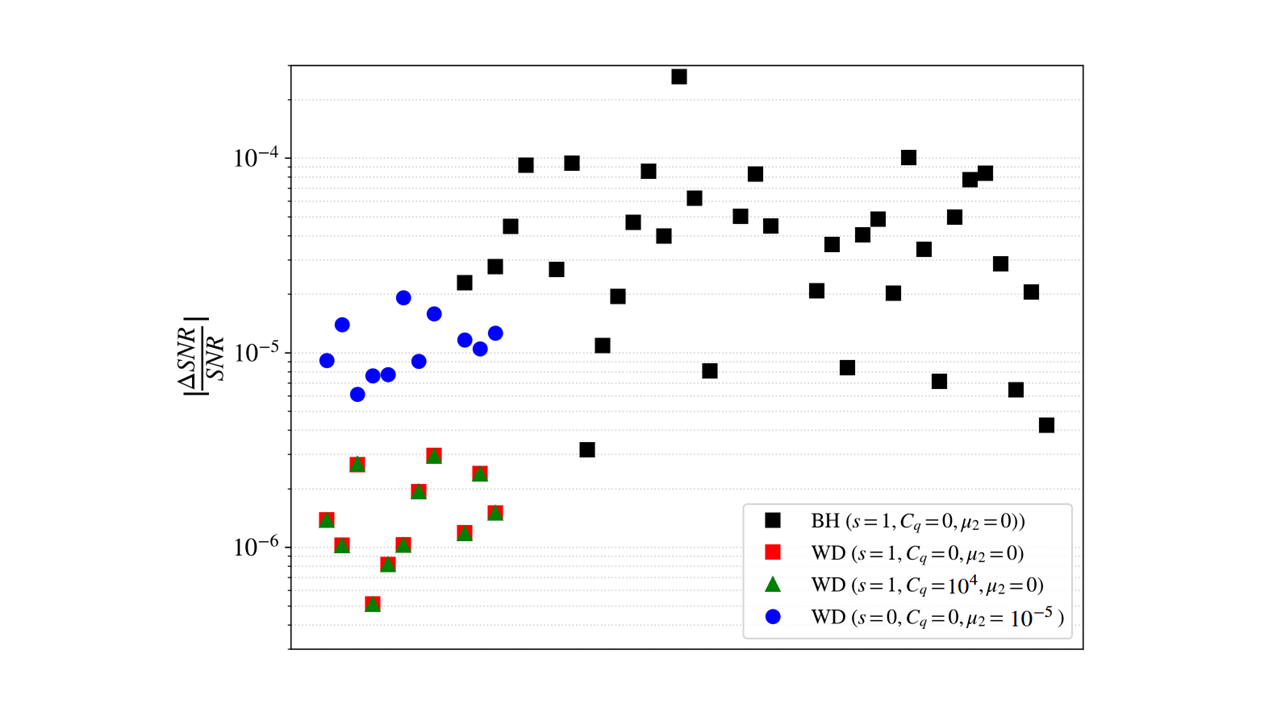}
\caption{The 
 relative difference of SNR ($\Delta$ SNR/SNR) between EMRIs consider or not  consider $\hat{s}$, $\mu_{2}$, and $C_{\rm Q}$. The black squares represent the EMRIs with $M_{\mathrm{co}}$ of tens of solar mass. The red squares, green triangles, and blue circles represent the EMRIs with $M_{\mathrm{co}}$ of one solar mass around.
\label{fig:snr}}
\end{figure}

In the frequency domain of Figure \ref{fig:wave}, the GW signals almost overlap, suggesting their SNRs are almost the same. We need to calculate the SNRs of GW signals to ensure that. The SNR of the signals can be defined as \cite{finn1992detection}
\begin{equation}
\label{SNR}
    \rho:=\sqrt{( h\vert h)}, 
\end{equation}
where $(h \vert h)$ is the inner product of signal $h(t)$ itself. The inner product between signal $a(t)$ and template $b(t)$ is as follows:
\begin{equation}
    (a\vert b)=2\int_{0}^{\infty}\frac{\tilde{a}^{*}(f)\tilde{b}(f)+{\tilde{a}(f)\tilde{b}^{*}(f)}}{S_{n}(f)}df,
\end{equation}
where $\tilde{a}(f)$ is the Fourier transform of $a(t)$, $\tilde{a}^{*}(f)$ is the complex conjugate of $\tilde{a}(f)$, and $S_{n}(f)$ is the power spectral density(PSD) of the GW detectors' noise. Throughout this paper, the PSD is taken to be the noise level of LISA. We first calculate the SNRs of several GW signals for EMRIs with $\hat{s} =0 $, $\mu_{2}=0$, and $C_{\rm Q}=0$, and then we calculate the SNRs of GW signals for EMRIs with different configurations of $\hat{s}$, $\mu_{2}$, and $C_{\rm Q}$.  Figure \ref{fig:snr} shows the relative difference between the former and latter signals' SNRs. We can see the influence of $\hat{s}$, $\mu_{2}$, and $C_{\rm Q}$ on the GW signal of EMRIs is about $10^{-4}$ to $10^{-6}$, so we may say the spin and quadrupole of the compact object are not important in constructing the gravitational waveform models of EMRIs.
\vspace{-9pt}
\vspace{-12pt}

In the time domain of Figure \ref{fig:wave}, during the last seconds of the orbit evolution,  we can see the influence of $\hat{s}$ and $\mu_{2}$ is significant, while the influence of $C_{\rm Q}$ is tiny. For further discussion of the quantified difference between the GW signals and the templates, we adopt the well-known matched-filtering technology. We use maximized fitting factor(overlap)
\begin{equation}
    \mathrm{FF}(a,b)=\underset{t_{s},\phi_{s}}{\mathbf{max}}\frac{(a(t)\vert b(t+t_{s})e^{i\phi_{s}})}{\sqrt{(a\vert a)(b\vert b)}},
\end{equation}
where $t_{s}$ is the time shift $t_{s}$ and $\phi_{s}$ is the phase shift. We use overlap to quantify the differences between the GW signals and the templates, and the results of overlap are calculated by PyCBC \cite{alex_nitz_2019_3247679}. 

\begin{figure}[H]
\includegraphics[width=11cm]{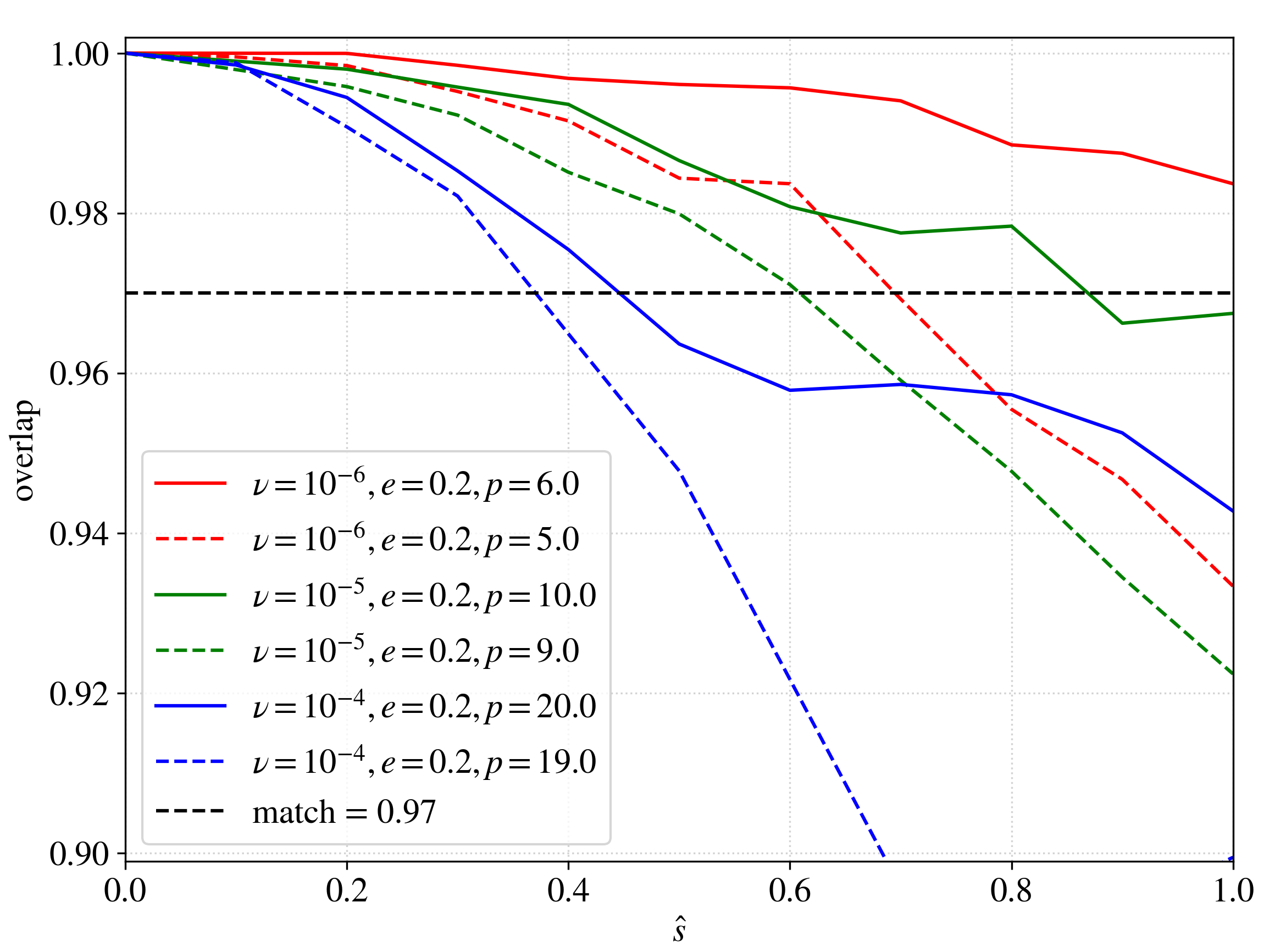}
\caption{Overlaps between the GW templates ($\hat{s}=0$) and the GW signals that changed with compact object's spin $\hat{s}$. $M_{\mathrm{bh}} = 10^{6} \mathrm{M_{\odot}}$, $a = 0.9$, $\iota_{0} = 0.0$, $\mu_{2} = 0$ and $C_{Q} = 0.0$. The red, green, and blue curves mark the mass ratio $10^{-6}$,  $10^{-5}$ and $10^{-5}$, respectively.
\label{fig:overlap_s}}
\end{figure}

\begin{figure}[H]
\includegraphics[width=12cm]{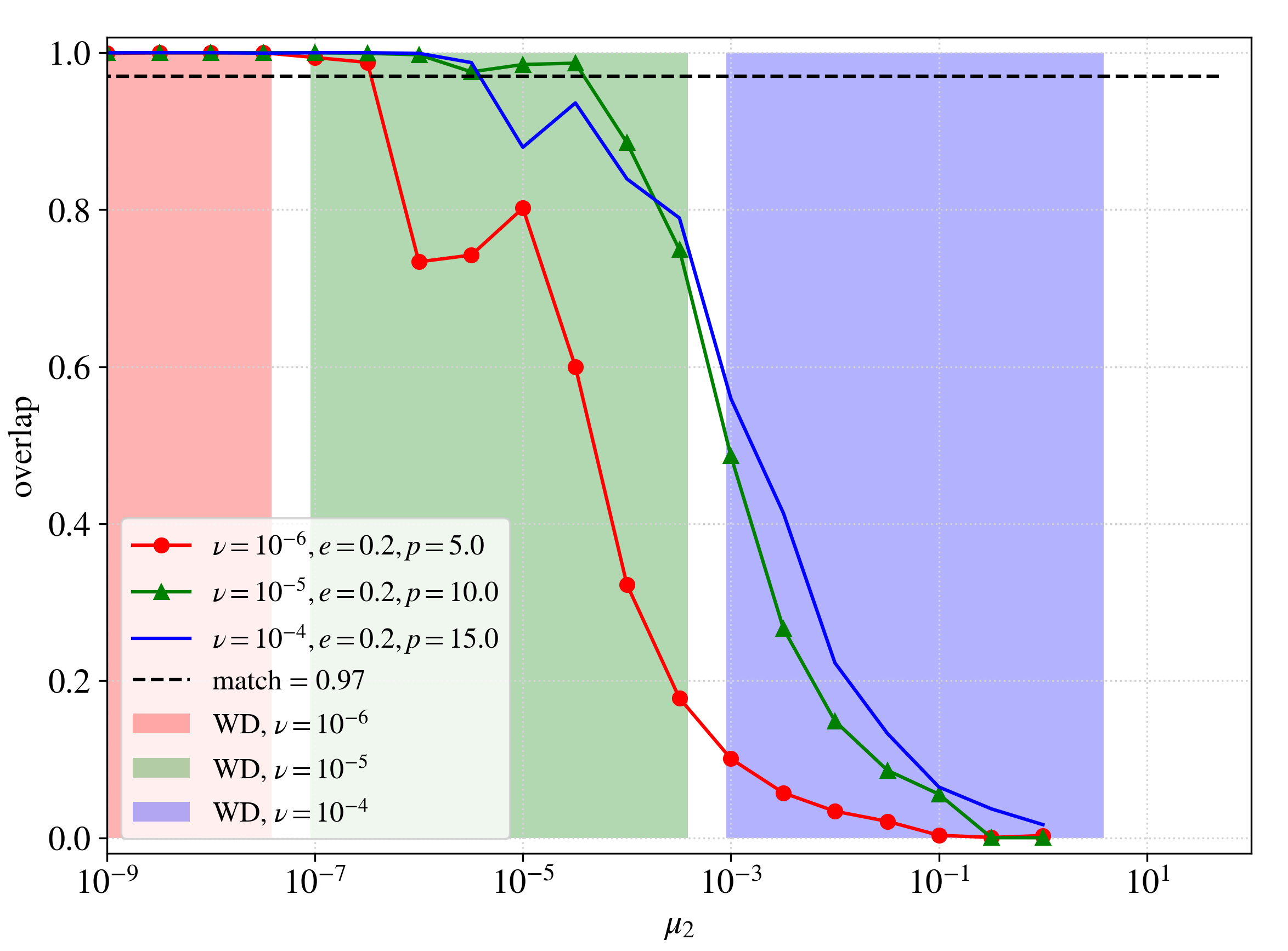}
\caption{Overlaps between the GW templates ($\mu_{2}=0$) and the GW signals that changed with compact object's tidal-induced quadrupole $\mu_{2}$. $M_{\mathrm{co}} = 1 \mathrm{M_{\odot}}$, $a = 0.9$, $\iota_{0} = 0.0$,  $\hat{s} = 0.0$ and $C_{Q} = 0.0$. The red, green, and blue curves mark the EMRIs with $\nu$ of $10^{-6}$,  $10^{-5}$, and $10^{-5}$, respectively. The red, green, and blue regions mark the rough range of $\mu_{2}$ for white dwarfs in EMRIs with different $\nu$.
\label{fig:overlap_mu2}}
\end{figure}

\begin{figure}[H]
\includegraphics[width=12cm]{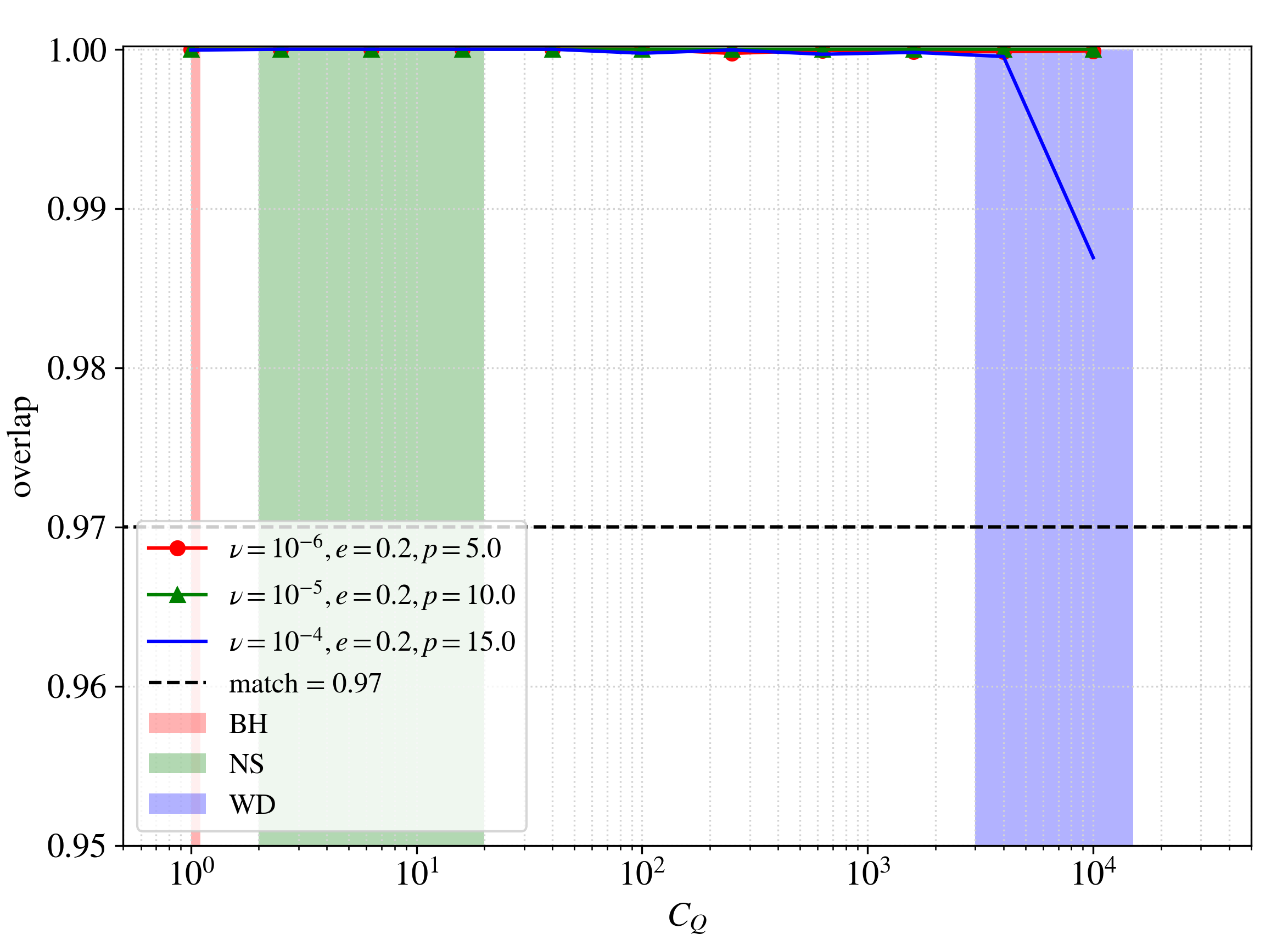}
\caption{Overlaps between the GW templates ($C_{Q} = 0$) and the GW signals that changed with compact object's spin-induced quadrupole $C_{\rm Q}$. $M_{\mathrm{co}} = 1 \mathrm{M_{\odot}}$, $a = 0.9$, $\iota_{0} = 0.0$, $\hat{s} = 1.0$ and $\mu_{2} = 0$. The red, green, and blue curves mark the EMRIs with $\nu$ of $10^{-6}$,  $10^{-5}$, and $10^{-4}$, respectively. The red, green, and blue regions mark the $C_{\rm Q}$ range for black holes, neutron stars, and white dwarfs.
\label{fig:overlap_cq}}
\end{figure}

The result of overlaps between the GW signals and templates are shown in \cref{fig:overlap_s,fig:overlap_mu2,fig:overlap_cq}. In this work, we set a criteria value of 0.97 for overlap as ref. \cite{yunes2011extreme}. If the overlap value is greater than 0.97, then we say the difference between the GW signals and templates is insignificant, and we could use this template in searching for GW signals of this kind. Figure \ref{fig:overlap_s} shows the overlaps between the GW templates ($\hat{s}=0$) and the GW signals that changed with the compact object's spin $\hat{s}$. We can see that for all GW signals, the overlaps decrease when $\hat{s}$ goes up. The overlaps down to 0.97 for most cases. Therefore, we may distinguish PBH ($\hat{s} \sim 0$) from compact objects with higher spin. Moreover, providing the information of redshift and eccentricity of the binary, there is a more systematic way for distinguishing PBH \cite{franciolini2022assess}. The tidal-induced quadrupole of the central black holes can be measured with great accuracy by LISA \cite{pani2019love, piovano2022extreme}. For compact objects, Figure \ref{fig:overlap_mu2} shows the overlaps between the GW templates ($\mu_{2}=0$) and the GW signals that changed with the compact object's tidal-induced quadrupole $\mu_{2}$.  Figure \ref{fig:overlap_mu2} shows that the overlaps decrease when $\mu_{2}$ goes up for all GW signals. However, for EMRIs of different $\nu$, only $\mu_{2}$ in the corresponding region is valid. There are only the $\mu_{2}$ ranges of white dwarfs shown in Figure \ref{fig:overlap_mu2}, and the $\mu_{2}$ ranges of black holes and neutron stars are too small to be shown. For the $\nu =10^{-6}$ case (red curve), the valid overlaps keep greater than 0.97, and we cannot identify $\mu_{2}$ of white dwarfs. For EMRIs with $\nu = 10^{-5}$ and $\nu = 10^{-4}$, especially $\nu = 10^{-4}$ case, it is possible to identify $\mu_{2}$ of white dwarfs. Figure \ref{fig:overlap_cq} shows the overlaps between the GW templates ($C_{\rm Q}=0$) and the GW signals that changed with the compact object's spin-induced quadrupole $C_{\rm Q}$. We can see that for most GW signals, the overlaps remain unchanged when $C_{\rm Q}$ goes up. In the case of $\nu = 10^{-4}$, the overlaps decrease a little for $C_{\rm Q} \sim 10^4$. Therefore, considering the results of \cref{fig:overlap_mu2,fig:overlap_cq}, for the compact object in the EMRIs with $\sim$1 $\rm M_{\odot}$ mass and $\nu$ of $10^{-4}$--$10^{-5}$, we could distinguish white dwarfs from other compact objects. Meanwhile, the quadrupole of black holes and neutron stars in EMRIs do not influence the GW signals.

\vspace{-9pt}
\unskip

\unskip

We use the Fisher information matrix to discuss further the parameter estimation accuracy for $m1$, $m2$, $a$, $\hat{s}$, $\mu_{2}$, $C_{Q}$, $\theta$, $\phi$, $D_{L}$. Additionally, we provide a visual representation of the posterior probability distributions and correlations among these parameters, as illustrated in \cref{fig:corner_mr5,fig:corner_mr4}.
 Fisher information matrix $\Gamma$ is an important method for parameter analysis and estimation \cite{cutler1994gravitational}. The matrix for a GW signal $h$ parameterized by $\lambda$ is given by
\begin{align}
    \Gamma_{i,j}=<\frac{\partial h}{\partial\lambda_{i}}|\frac{\partial h}{\partial\lambda_{j}}>,
\end{align}
where $\lambda=(m1, m2, a, \hat{s}, \mu_{2}, C_{Q}, \theta, \phi, D_{L})$ for EMRIs. The parameter estimation errors $\Delta\lambda$ due to Gaussian noise have the normal distribution $\mathcal{N}(0,\Gamma^{-1})$ in the case of high SNR, and the root-mean-square errors in the general case can be approximated as
\begin{align}
    \Delta\lambda_{i}=\sqrt{(\Gamma^{-1})_{i,i}}.
\end{align}

{To 
 assess the parameter estimation precision, we calculate the relative errors $(\Delta \lambda / \lambda)$ for $m_1$, $m_2$, $a$, $\hat{s}$, $\mu_2$, $C_Q$, $\theta$, $\phi$, and $D_L$. As shown in Table \ref{tab:table3}, the accuracy improves as the mass ratio approaches the intermediate range, $\nu = 10^{-4}$$\sim$$10^{-5}$.} For the $\nu =10^{-6}$ case, we cannot obtain valid results of $\Delta \hat{s}/\hat{s}$, $\Delta \mu_{2} / \mu_{2}$ and $\Delta C_{\rm Q} / C_{\rm Q} $ since they are greater than one. For the $\nu =10^{-5}$ case, $\Delta \hat{s}/\hat{s}$ achieves the accuracy  of $\sim$$10^{-1}$, and $\Delta \mu_{2} / \mu_{2}$ obtains the accuracy of $\sim$$10^{-1}$. For the $\nu =10^{-4}$ case, $\Delta \hat{s}/\hat{s}$ achieves the accuracy of $\sim$$10^{-2}$, $\Delta \mu_{2} / \mu_{2}$  reaches the accuracy of $\sim$$10^{-1}$, and similarly, even though in this case we have a larger mass ratio, we still unable to obtain the valid accuracy of $ \Delta C_{\rm Q} / C_{\rm Q}$ at those waveforms with $ \hat{s}=0.8 $.

In particular, according to Table \ref{tab:table2}, we can see that since the spin of WDs can reach up to $10$, we made another evaluation on the parameter estimation for WDs with higher spin values, and the results are shown in Table \ref{tab:table4}. It can be noted that the higher spin is positively influential in improving the accuracy of the evaluation on the parameter estimation, and in comparison to $\hat{s}=0.8$, the spin-induced quadrupole $C_{\rm Q}$ also shows a sizable overlap, with the accuracy even reached $10^{-2}$ at the $\nu = 10^{-4}$ case. However, it is worth mentioning that the spin-induced quadrupole $C_{\rm Q}$, in general, does not show any measurable overlaps, although it probably works in the situation of higher spins and higher mass rations, which means that it has a negligible impact on our method of distinguishing compact objects.

\begin{table}[H] 
\caption{Fisher Matrix Results.
\label{tab:table3}}
\begin{adjustwidth}{-\extralength}{0cm}
\newcolumntype{C}{>{\centering\arraybackslash}X}
\begin{tabularx}{\fulllength}{CCCCCCCC}
\toprule
\boldmath{$\nu$} & \boldmath{$\Delta m_{1} / m_{1}$} & \boldmath{$\Delta m_{2} / m_{2}$} & \boldmath{$\Delta a / a $} & \boldmath{$\Delta \hat{s} / \hat{s}$} &  \boldmath{$\Delta\mu_{2} / \mu_{2} $}  & \boldmath{$\Delta C_{Q} / C_{Q}$} & \boldmath{$\Delta D_{L} / D_{L}$}\\
\midrule
$10^{-4}$ & $9.20 \times 10^{-6}$ & $8.85 \times 10^{-4}$ & $9.27 \times 10^{-5}$ & $4.69 \times 10^{-2}$  & $1.49 \times 10^{-1}$ & $-$ & $1.44 \times 10^{-2}$ \\

$10^{-5}$ & $6.28 \times 10^{-5}$ & $6.39 \times 10^{-4}$ & $6.69 \times 10^{-5}$ & $6.08 \times 10^{-1}$  & $2.90 \times 10^{-1}$ & $-$ & $7.16 \times 10^{-2}$\\

$10^{-6}$ & $7.10 \times 10^{-5}$ & $7.71 \times 10^{-4}$ & $8.31 \times 10^{-5}$ & $-$  & $-$ & $-$  & $6.92 \times 10^{-2}$\\
\bottomrule
\end{tabularx}
\end{adjustwidth}
\end{table}
\vspace{-12pt}
\begin{table}[H] 
\caption{Fisher Matrix Results for White Dwarfs.
\label{tab:table4}}
\begin{adjustwidth}{-\extralength}{0cm}
\newcolumntype{C}{>{\centering\arraybackslash}X}
\begin{tabularx}{\fulllength}{CCCCCCCC}
\toprule
\boldmath{$\nu$} & \boldmath{$\Delta m_{1} / m_{1}$} & \boldmath{$\Delta m_{2} / m_{2}$} & \boldmath{$\Delta a / a $} & \boldmath{$\Delta \hat{s} / \hat{s}$} &  \boldmath{$\Delta\mu_{2} / \mu_{2} $ } & \boldmath{$\Delta C_{Q} / C_{Q}$} & \boldmath{$\Delta D_{L} / D_{L}$}\\
\midrule
$10^{-4}$ & $2.64 \times 10^{-5}$ & $2.68 \times 10^{-4}$ & $2.85 \times 10^{-5}$ & $3.40 \times 10^{-2}$  & $6.13 \times 10^{-3}$ & $7.71 \times 10^{-2}$ & $3.69 \times 10^{-2}$ \\

$10^{-5}$ & $6.27 \times 10^{-5}$ & $6.41 \times 10^{-4}$ & $6.68 \times 10^{-5}$ & $4.03 \times 10^{-2}$  & $3.18 \times 10^{-1}$ & $7.24 \times 10^{-1}$ & $7.30 \times 10^{-2}$\\

$10^{-6}$ & $1.20 \times 10^{-5}$ & $1.21 \times 10^{-4}$ & $1.19 \times 10^{-5}$ & $1.15 \times 10^{-2}$  & $-$ & $-$  & $1.21 \times 10^{-2}$\\
\bottomrule
\end{tabularx}
\end{adjustwidth}
\end{table}

\begin{figure}[H]
\begin{adjustwidth}{-\extralength}{0cm}
\centering
\includegraphics[width=18.5cm]{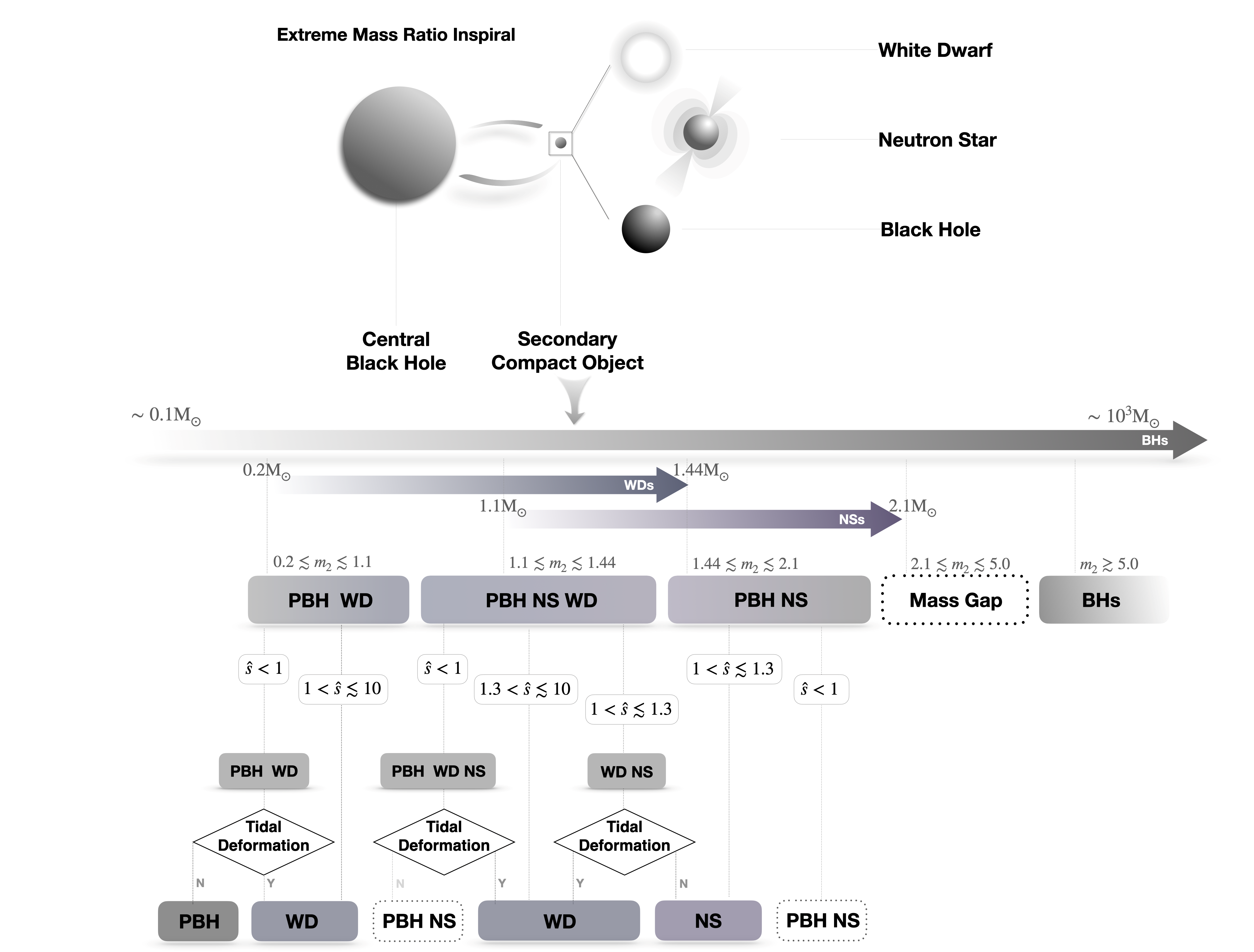}
\end{adjustwidth}
\caption{The flowchart for our method of distinguishing Compact Objects.
\label{fig:flow}}
\end{figure}

In order to provide a lucid explanation, we have produced a flowchart to describe how we go about our method of distinguishing compact objects. Since our primary concern is how to distinguish WDs and NSs from PBHs, rather than SOBHs from PBHs, we only consider the mass range from $0.2 \rm M_{\odot}$ to $2.1 \rm M_{\odot}$. Based on Table \ref{tab:table2} and the results of Fisher Information Matrix exhibited above, our method can be divided into three steps which rely mainly on three physical characteristics $m_{2}$, $\hat{s}$, $C_{\rm Q}$, respectively, where $C_{\rm Q}$ is not taken into account as it does not have a significant effect on our waveforms. In the first step, we roughly divide them into three parts corresponding to their respective mass ranges. The second step is crucial because we use the different spin ranges to further constrain them, which enables us to directly distinguish WDs from other compact objects when their spins exceed $1.3$, and even NSs from PBHs when $1.0<\hat{s}\lesssim1.3$. To confirm the identity of an object, the decisive step is that {whether} it exhibits an observable tidal effect, which requires $\mu_{2}\gtrsim 10^{-9}$, as indicated by the results of Fisher Information Matrix in Table \ref{tab:table4}.
Figure \ref{fig:flow} shows that the vast majority of scenarios can be distinguished successfully, except in one scenario where $\hat{s}<1.0$ and both NSs and PBHs have negligible tidal effects, making it hard to separate them. Nevertheless, when the spins of NSs are larger than those of PBHs, they can be distinguished even when their spins are both below $1.0$. This is consistent with our result of the Bayes factor, which is given in Table \ref{tab:table5}.

\begin{figure}[H]
\begin{adjustwidth}{-\extralength}{0cm}
\centering
\includegraphics[width=18cm]{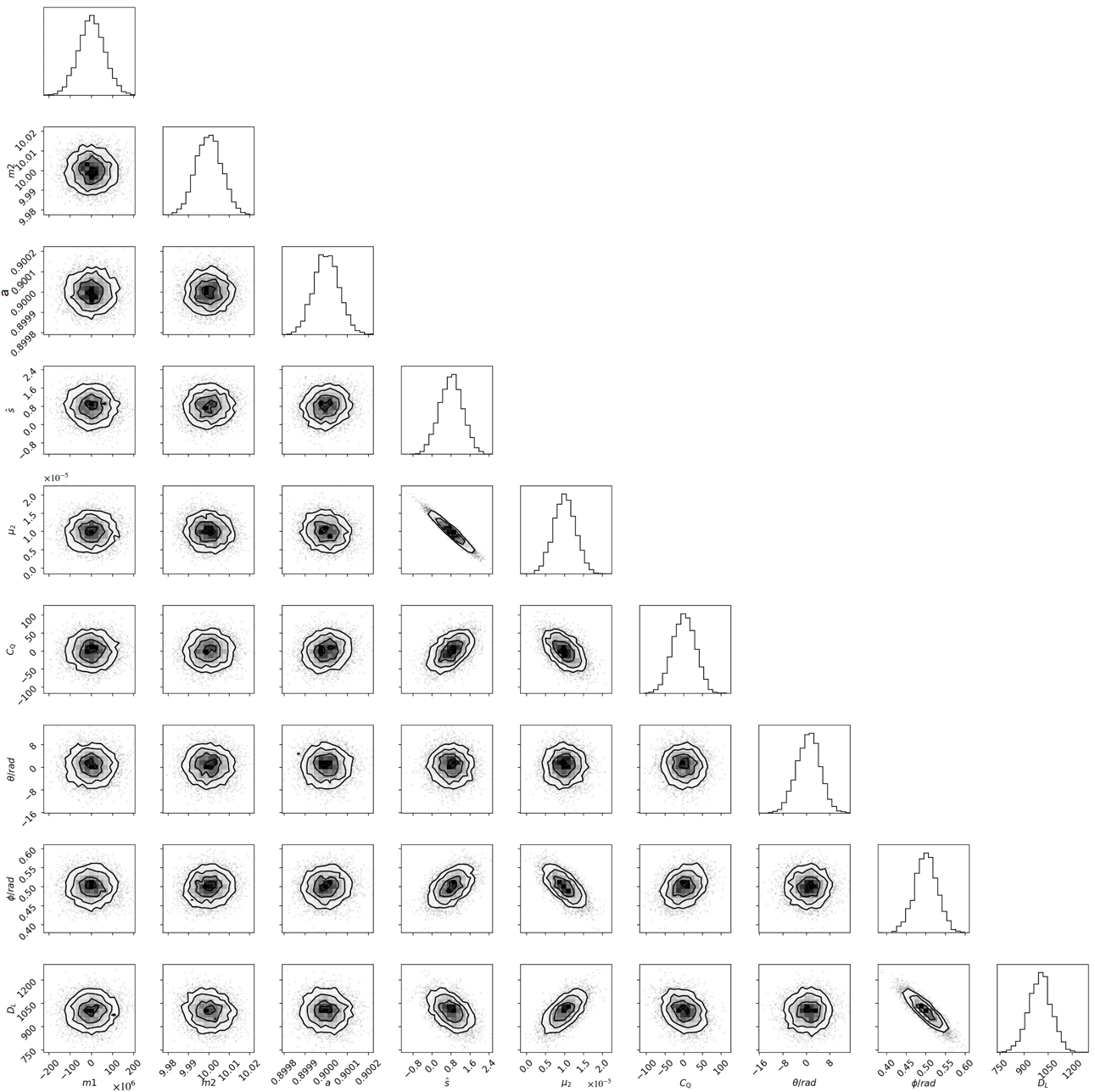}
\end{adjustwidth}
\caption{{For 
 the case of $\nu = 10^{-5}$, the corner 
 plot illustrates the probability distributions and correlations among the parameters $m_1$, $m_2$, $a$, $\hat{s}$, $\mu_2$, $C_Q$, $\theta$, $\phi$, and $D_L$. The diagonal panels display the marginalized one-dimensional posterior probability distributions for each parameter, while the off-diagonal panels represent the two-dimensional marginalized posterior distributions for parameter pairs. The contours denote the 11.8\%, 39.3\%, 67.5\%, and 86.4\% credible intervals.}
\label{fig:corner_mr5}}
\end{figure}   
\unskip

\begin{figure}[H]
\begin{adjustwidth}{-\extralength}{0cm}
\centering
\includegraphics[width=18cm]{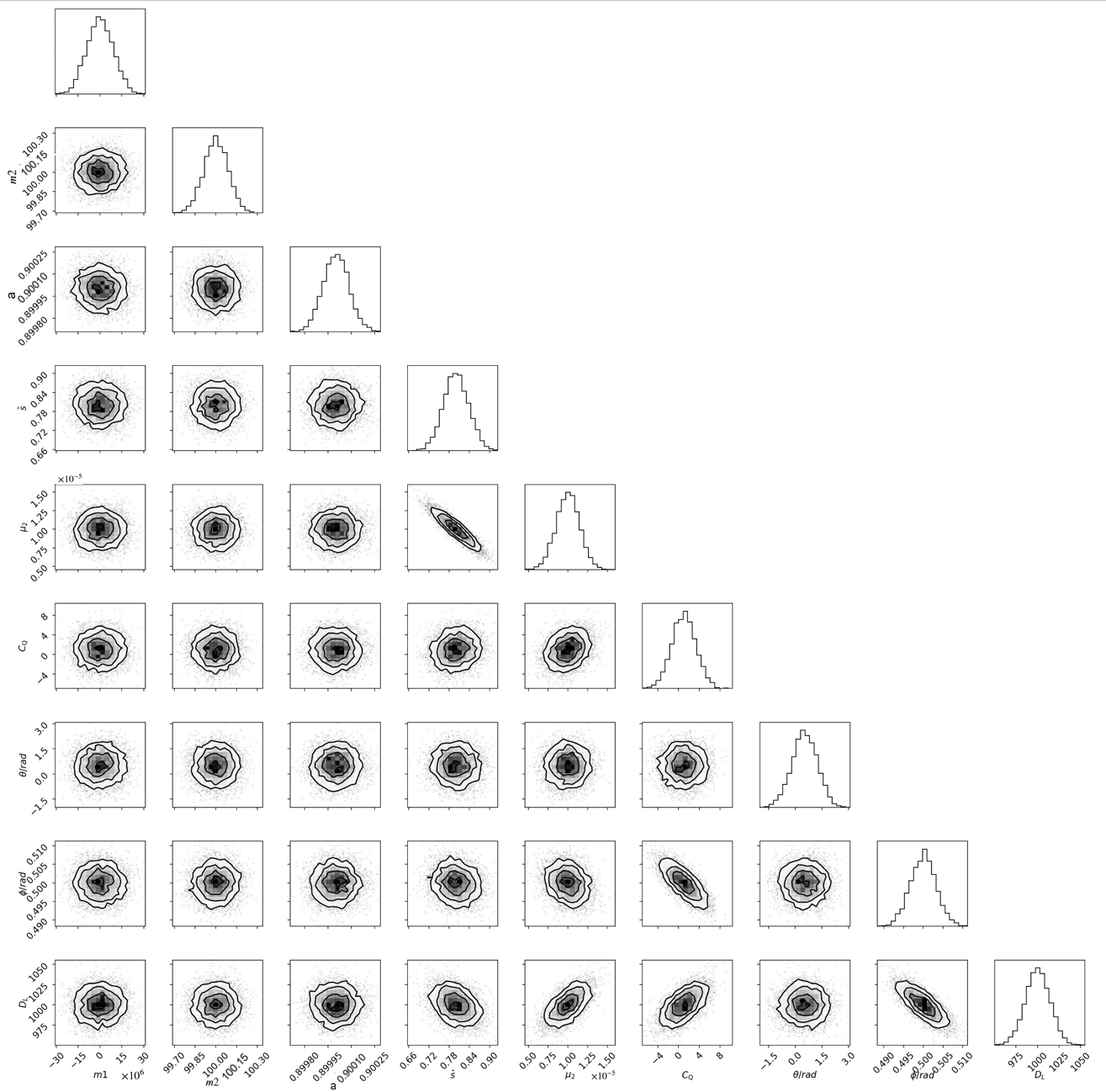}
\end{adjustwidth}
\caption{{For 
 the case of $\nu = 10^{-4}$, the corner 
 plot illustrates the probability distributions and correlations among the parameters $m_1$, $m_2$, $a$, $\hat{s}$, $\mu_2$, $C_Q$, $\theta$, $\phi$, and $D_L$. The diagonal panels display the marginalized one-dimensional posterior probability distributions for each parameter, while the off-diagonal panels represent the two-dimensional marginalized posterior distributions for parameter pairs. The contours denote the 11.8\%, 39.3\%, 67.5\%, and 86.4\% credible intervals.}
\label{fig:corner_mr4}}
\end{figure}   

To quantify the extent to which our waveforms show that WDs and NSs can be distinguished from BHs, we introduce a simple linearized equation of the Bayes factor from Moore et al. \cite{moore2021tests}:
\begin{equation}
    \ln \mathcal{B}=\ln \bigg(\frac{\Pi}{A}\sqrt{\frac{2\pi}{\rho}}\bigg)+\frac{(z+\rho\sqrt{2(1-\mathcal{M})}\cos{\iota})^{2}}{2},
\end{equation}
where $\Pi$ is the prior odds ratio, $A = \alpha_{max}-\alpha_{min}$ is the prior range on $\alpha$, $\iota$ is defined as the angle between signals, typically we consider setting $\cos{\iota}=\pm 1$ and it is a random choice of sign associated with model error $\Delta \hat{h}$. Similarly, $z\sim \mathcal{N}(0,1)$ is a random number associated with the noise realization. For mismatch with the small angle approximation, we have $1-\mathcal{M}\approx \frac{|\Delta \hat{h}|^{2}}{2}$.
Based on Ye Jiang's work \cite{jiang2023resolving}, we set $\Pi=A=1$, $z=0$, $\cos{\iota}=1$ and finally obtain the simplified expression of Bayes factor:
\begin{align}
    \ln \mathcal{B}=\ln \left(\frac{\sqrt{2\pi}}{\rho}
    \right )+\rho^{2}(1-\mathcal{M}).
\end{align}
\vspace{-12pt}
\begin{table}[H] 
\caption{Bayes factors ($\ln \mathcal{B}$) between the waveforms for PBH and waveforms with different $\hat{s}, \mu_{2}, C_{Q}$ for NS and WD.
\label{tab:table5}}
\begin{adjustwidth}{-\extralength}{0cm}
\newcolumntype{C}{>{\centering\arraybackslash}X}
\begin{tabularx}{\fulllength}{CCCCCCCC}
\toprule
& \boldmath{$\nu$} & \boldmath{$\hat{s}$} & \boldmath{$\mu_{2}$} & \boldmath{$C_{Q}$} &  \boldmath{$\mathrm{SNR}$}  & \boldmath{$ 1-\mathcal{M} $} & \boldmath{$\ln \mathcal{B} $}\\

\midrule
$\mathrm{NS}$ & $10^{-4}$ & $0.8$ & $0.0$ & $2.0$  & $91.9$ & $0 $ & $-1.3 $ \textsuperscript{a} \\
$\mathrm{NS}$ & $10^{-4}$ & $0.8$ & $0.0$ & $10.0$  & $91.9$ & $0 $ & $-2.7 $ \textsuperscript{b} \\
$\mathrm{NS}$ & $10^{-4}$ & $1.3$ & $0.0$ & $10.0$  & $91.9$ & $0.128 $ & $1079.6 $ \\
$\mathrm{NS}$ & $10^{-5}$ & $0.8$ & $0.0$ & $10.0$  & $31.8$ & $0 $ & $-2.5 $ \textsuperscript{c} \\
$\mathrm{NS}$ & $10^{-5}$ & $1.3$ & $0.0$ & $10.0$  & $31.8$ & $0.009 $ & $6.5 $ \\
$\mathrm{NS}$ & $10^{-6}$ & $0.8$ & $0.0$ & $10.0$  & $16.6$ & $0 $ & $-1.9 $ \textsuperscript{d} \\
$\mathrm{NS}$ & $10^{-6}$ & $1.3$ & $0.0$ & $10.0$  & $16.6$ & $0.012 $ & $1.4$ \\

$\mathrm{WD}$ & $10^{-4}$ & $0.2$ & $1.0\times 10^{-3}$ & $1.0\times 10^{4}$  & $91.7$ & $0.589 $ & $4950.9$ \\
$\mathrm{WD}$ & $10^{-4}$ & $0.4$ & $1.0\times 10^{-3}$ & $1.0\times 10^{4}$  & $91.7$ & $0.603 $ & $5066.7$ \\
$\mathrm{WD}$ & $10^{-4}$ & $0.8$ & $1.0\times 10^{-3}$ & $1.0\times 10^{4}$  & $91.7$ & $0.591 $ & $4968.5$ \\
$\mathrm{WD}$ & $10^{-4}$ & $0.8$ & $0.0$ $^\dagger$ & $1.0\times 10^{4}$ & $91.9$ & $0 $ & $-2.7$ \textsuperscript{e} \\
$\mathrm{WD}$ & $10^{-4}$ & $1.0$ & $1.0\times 10^{-3}$ & $1.0\times 10^{4}$  & $91.7$ & $0.593 $ & $4978.2$ \\
$\mathrm{WD}$ & $10^{-4}$ & $1.3$ & $1.0\times 10^{-3}$ & $1.0\times 10^{4}$  & $91.7$ & $0.591 $ & $4965.4$ \\
$\mathrm{WD}$ & $10^{-4}$ & $1.8$ & $1.0\times 10^{-3}$ & $1.0\times 10^{4}$  & $91.7$ & $0.603 $ & $5071.1$ \\
$\mathrm{WD}$ & $10^{-4}$ & $4.0$ & $1.0\times 10^{-3}$ & $1.0\times 10^{4}$  & $91.7$ & $0.600$ & $5039.9$ \\
$\mathrm{WD}$ & $10^{-5}$ & $0.4$ & $1.0\times 10^{-5}$ & $1.0\times 10^{4}$  & $31.8$ & $0.004 $ & $2.3$ \\
$\mathrm{WD}$ & $10^{-5}$ & $0.8$ & $1.0\times 10^{-5}$ & $1.0\times 10^{4}$  & $31.8$ & $0.020 $ & $17.6$ \\
$\mathrm{WD} $ & $10^{-5}$ & $1.3$ & $1.0\times 10^{-5}$ & $1.0\times 10^{4}$  & $31.8$ & $0.034 $ & $32.0$ \\
$\mathrm{WD} $ & $10^{-5}$ & $2.7$ & $1.0\times 10^{-5}$ & $1.0\times 10^{4}$  & $31.8$ & $0.055 $ & $53.0$ \\
$\mathrm{WD}$ & $10^{-6}$ & $0.8$ & $1.0\times 10^{-9}$ & $1.0\times 10^{4}$  & $16.6$ & $0.141 $ & $36.9$ \\
$\mathrm{WD}$ & $10^{-6}$ & $1.3$ & $1.0\times 10^{-9}$ & $1.0\times 10^{4}$  & $16.6$ & $0.005 $ & $-0.6$ \textsuperscript{f}\\
$\mathrm{WD}$ & $10^{-6}$ & $3.6$ & $1.0\times 10^{-9}$ & $1.0\times 10^{4}$  & $16.6$ & $0.040 $ & $9.1$ \\
\bottomrule
\end{tabularx}
\end{adjustwidth}
\noindent{\footnotesize{
\textsuperscript{a}, \textsuperscript{b} When 
 these two NSs in $\nu=10^{-4}$ compared with PBHs ($\hat{s} = 0$, $\mu_{2} = 0, C_{\rm Q} = 1$),  both $\ln \mathcal{B} \sim 9.0 \times  10^{2}$. 
\textsuperscript{c} When this NS in $\nu=10^{-5}$ compared with PBHs ($\hat{s} = 0$, $\mu_{2} = 0$, $C_{\rm Q} = 1$), $\ln \mathcal{B} = 35.7$. 
\textsuperscript{d} When this NS in $\nu=10^{-6}$ compared with PBHs ($\hat{s} = 0$, $\mu_{2} = 0$, $C_{\rm Q} = 1$), $\ln \mathcal{B} = 10.3$. 
\textsuperscript{e}, $^\dagger$ When this WD in $\nu=10^{-4}$ compared with PBHs ($\hat{s} = 0$, $\mu_{2} = 0$, $C_{\rm Q} = 1$), $\ln \mathcal{B} \sim 1.0 \times 10^{3}$. The purpose of setting $\mu_{2} = 0$ here is to contrast with other sets of WDs and, therefore, validate the crucial role of tidal deformation in distinguishing COs from PBHs. Under normal circumstances, the tidal deformation of WDs is observable.
\textsuperscript{f} When this WD in $\nu=10^{-6}$ compared with PBHs ($\hat{s} = 0$, $\mu_{2} = 0$, $C_{\rm Q} = 1$), $\ln \mathcal{B} = 27.1$. }}
\end{table}

To validate the effectiveness of our distinguishing mechanism and quantify the differences among waveform data, we calculated and compared the Bayes factors of all possible COs waveforms with three types of PBHs waveform templates ($\hat{s} = 0$, $\mu_{2} = 0, C_{\rm Q} = 1$), ($\hat{s} = 0.8$, $\mu_{2} = 0, C_{\rm Q} = 1$) and ($\hat{s} = 1.0$, $\mu_{2} = 0, C_{\rm Q} = 1$), as displayed in Table \ref{tab:table5}. Following the workflow depicted in the previous diagram, our distinguishing mechanism primarily involves segregating by the mass $m_{2}$, then distinguishing various spin ranges, and subsequently assessing whether there is tidal deformation to discern different types of COs, where the corresponding parameters are $\hat{s}$, $\mu_{2}$, $C_{\rm Q}$. During the calculation of $\ln \mathcal{B}$, to understand more precisely the effect of these three parameters $\hat{s}$, $\mu_{2}$ and $C_{\rm Q}$ on the extent to which waveforms of Compact Objects can be distinguished from templates of PBHs, we set different values for $\hat{s}$, $\mu_{2}$ and $C_{\rm Q}$ of NSs and WDs according to their physical characteristics, and meanwhile set parallel values for different mass ratios, which allowed us to distinguish WDs and NSs from PBHs under different SNRs.

Therefore, we listed 22 waveforms of COs, comparing them with both PBHs without spin and those with different spins. It is noteworthy that when the spin of COs is $\hat{s} \lesssim 0.8$, we compared it with PBHs with $\hat{s} = 0.8$, whereas when the spin of COs exceeds $1.0$, a comparison was made with PBHs with $\hat{s} = 1.0$. Table \ref{tab:table5} presents $\ln \mathcal{B}$ results compared to PBHs templates with $\hat{s} = 0.8$ or $\hat{s} = 1.0$, as well as several key data points highlighted, indicating extra significant findings.

Our data reveal that effective differentiation primarily depends on whether COs exhibit significant tidal deformation ($ \mu_{2} \gtrsim 10^{-9}$), followed by substantial spin ($ \hat{s} \gtrsim 0.8$). This implies that even with considerable spin, distinguishing between COs and PBHs with large spin becomes challenging in the absence of tidal effects, as demonstrated by the data row WD ($\hat{s} = 0.8$, $\mu_{2} = 0, C_{\rm Q} = 3.0\times 10^{3}$) in $\nu=10^{-4}$ v.s. other WDs data. This also elucidates the difficulty in distinguishing NSs from the PBHs with large spin, which is evident in all the NS waveforms with $\hat{s} = 0.8$ and $\mu_{2} = 0$. Furthermore, the negligible impact of spin-induced quadrupole $C_{\rm Q}$ on the waveforms differentiation aligns with our Fisher information matrix results, further validating the reliability and efficacy of our distinguishing mechanism. Additionally, we observed a significant increase in $\ln \mathcal{B}$ in scenarios with higher mass ratios ($\nu = 10^{-4}$ and $\nu = 10^{-5}$), indicating that there is a certain positive correlation between $\ln \mathcal{B}$ and mass ratio, and our distinguishing effectiveness becomes more pronounced at a mass ratio of $\nu = 10^{-4}$.

Based on our distinguishing mechanism and the supporting data presented in Table \ref{tab:table5}, our primary conclusions are as follows: our waveform templates can effectively distinguish between WDs and PBHs; NSs and PBHs can also be differentiated under certain conditions, particularly when the spin of PBHs is lower than that of NSs. This encompasses two scenarios: (1) when the PBHs spin is zero ($\hat{s} = 0$, $\mu_{2} = 0, C_{\rm Q} = 1$), as demonstrated in the details of the annotation of the seven NSs data; and (2) when the PBHs have spin, but the NSs have spin exceeding $1.0$, as demonstrated by the three sets of NSs data with $\hat{s} = 1.3$.

\section{Conclusions}
\label{conclusion}

In the current work, we employ the MPD equation as an alternative to the test particle approximation to solve the orbits of the compact object and then simulate GW signals from EMRIs that include the spin and quadrupoles of the compact object. In this instance, the compact object in the EMRIs is characterized as an extended body, as opposed to a test particle.

In order to investigate the influence of spin and quadrupoles of compact objects on GW signals and explore the potential for revealing the structure of compact objects in extreme mass ratio inspirals, we consider the case where the mass of the compact object is approximately one solar mass. We examine three scenarios where the compact object could be a primordial black hole (PBH), a neutron star (NS), or a white dwarf (WD).

Initially, for all potential compact object types, we evaluate the accuracy of parameter estimation for EMRIs over a mass ratio range of $10^{-6}$ to $10^{-4}$, employing the Fisher Information Matrix (FIM). The results are summarized in Table \ref{tab:table3}. Due to the wide range of spin values exhibited by WDs, we present the FIM results for high spin values separately in Table \ref{tab:table4}. In general, the results indicate that the precision for spin estimation can reach approximately $10^{-2}$, while the accuracy for $\mu_2$ can reach $10^{-1}$ for mass ratios between $10^{-4}$ and $10^{-5}$. Notably, in scenarios with higher spin values, the detection accuracy for both $C_{q}$ and $\mu_2$ significantly improves, reaching a precision of $10^{-2}$ at a mass ratio of $10^{-4}$ and $10^{-1}$ at a mass ratio of $10^{-5}$. This suggests that higher spin values enhance the detectability of the quadrupole effects, including both tidal-induced and spin-induced quadrupoles. Additionally, we present the probability distributions and correlations for all estimated parameters.

Subsequently, for each scenario, we calculate the overlap between the gravitational wave (GW) signals generated with varying spin values ($\hat{s}$), spin-induced quadrupoles ($C_{q}$), and tidal-induced quadrupoles ($\mu_2$), and templates with $\hat{s}$, $C_{q}$, and $\mu_2$ set to zero. The overlap results are depicted in \cref{fig:overlap_s,fig:overlap_mu2,fig:overlap_cq}, from which we obtain the following findings: the spin of all types of compact objects induces detectable variations in the GW signals, particularly for mass ratios in the range of $10^{-5}$ to $10^{-4}$, aligning with previous studies \cite{Nitz_spin_2013}. The tidal-induced quadrupoles only influence the GW signals when the compact object is a WD, especially in cases where the mass ratio approaches $\nu = 10^{-4}$. Spin-induced quadrupoles, on the other hand, have a negligible effect on EMRI waveforms. The overlap analysis reveals that spin and tidal-induced quadrupoles serve as key factors in differentiating PBHs from WDs and NSs, forming the basis of our distinguishing mechanism.

Accordingly, based on the extended body model and the distinguishing mechanism involving the parameters $\hat{s}$, $\mu_2$, and $C_{\rm Q}$, we calculate the Bayes factor for 22 different compact objects (COs), including WDs and NSs, and compare these results with various PBH cases ($\hat{s} = 0$, $\mu_2 = 0, C_{\rm Q} = 1$), ($\hat{s} = 0.8$, $\mu_2 = 0, C_{\rm Q} = 1$), and ($\hat{s} = 1.0$, \mbox{$\mu_2 = 0, C_{\rm Q} = 1$}). Our results indicate that detectable tidal deformations lead to higher Bayes factors, highlighting the crucial role of $\mu_2$ in distinguishing compact objects. While spin also plays a role, particularly in the absence of tidal effects, spin alone can be effective in distinguishing COs from PBHs. Consequently, PBHs can be reliably distinguished from WDs due to the strong tidal effects inherent to WDs, regardless of whether PBHs possess spin. Similarly, NSs can also be differentiated from PBHs in cases where the NS spin exceeds that of PBHs.

{This study primarily investigates the distinction between PBHs, WDs, and NSs. However, it is worth noting that many other types of compact objects, such as exotic compact objects (ECOs) and other BH mimickers, have not been considered here. The works presented in \cite{Maselli_2018,Addazi_2019} provide valuable insights into distinguishing black holes (BHs) from exotic compact objects (ECOs) by employing tidal heating and tidal deformability as key mechanisms. Their results demonstrate that tidal heating is absent in ECOs due to the lack of a horizon, whereas BHs exhibit nonzero tidal heating. Additionally, the tidal Love numbers (TLNs) for BHs are exactly zero, while those for ECOs are small but finite.}

{Building upon these studies, future work could focus on developing similar techniques to distinguish PBHs from ECOs and other BH mimickers inspired by quantum gravity models. Such efforts would contribute to a deeper understanding of the diverse compact object populations and their fundamental physical characteristics, further advancing the fields of gravitational wave detection and multi-messenger astrophysics.}

\vspace{6pt}

\authorcontributions{Conceptualization, W.-B.H.; methodology, S.-C.Y. and W.-B.H.; software, S.-C.Y. and X.-Y.Z.; validation, L.-J.X., S.-C.Y. and X.-Y.Z.; formal analysis, L.-J.X. and S.-C.Y.; investigation, L.-J.X., S.-C.Y. and R.-D.T.; resources, W.-B.H.; data curation, L.-J.X. and S.-C.Y.; writing---original draft preparation, L.-J.X., S.-C.Y., R.-D.T. and Y.-H.Z.; writing---review and editing, L.-J.X., S.-C.Y. and W.-B.H.; visualization, L.-J.X. and S.-C.Y.; supervision, S.-C.Y. and W.-B.H.; project administration, W.-B.H.; funding acquisition, W.-B.H. All authors have read and agreed to the published version of the manuscript.}

\funding{This work is supported by The National Key R\&D Program of China (No. 2021YFC2203002), NSFC No. 11773059, No. 12173071, and No. 12473075. This work made use of the High-Performance Computing Resource in the Core Facility for Advanced Research Computing at Shanghai Astronomical Observatory.}

\institutionalreview{Not applicable.}

\informedconsent{Not applicable.}

\dataavailability{The data sets generated during the current study are available from the corresponding author upon reasonable request.}




\acknowledgments{We thank Carlos 
 A. Benavides-Gallego, Ahmadjon Abdujabbarov and Imene Belahcene for their valuable advice on this work.}

\conflictsofinterest{The authors declare no conflicts of interest.} 



\abbreviations{Abbreviations}{
The following abbreviations are used in this manuscript:\\

\noindent 
\begin{tabular}{@{}ll}
EMRI & Extreme mass-ratio inspirals\\
GW & Gravitational wave\\
FIM & Fisher information matrix\\
SNR & signal-to-noise ratio\\
MBH & Massive black hole\\
LISA & Laser Interferometer Space Antenna\\
MPD  &  {Mathisson-Papapetrou}-Dixon\\
ISCO & Innermost stable circular orbit\\
PBH  & Primordial black hole\\
NS & Neutron star\\
WD & White dwarf\\
EOS  & Equation of state\\
\end{tabular}
}




\begin{adjustwidth}{-\extralength}{0cm}

\reftitle{References}

\PublishersNote{}
\end{adjustwidth}
\end{document}